\documentclass[12pt]{article}

\pdfoutput=1

\usepackage{fullpage}
\usepackage{graphicx}
\usepackage[english]{babel}
\usepackage{color}
\usepackage[usenames]{xcolor}
\usepackage{amsfonts}
\usepackage{indentfirst}
\usepackage{setspace}
\usepackage{verbatim}
\usepackage{upgreek}
\usepackage{amsmath}
\usepackage{amsthm}
\usepackage{amssymb}
\usepackage[normalem]{ulem}
\usepackage{float}
\usepackage{latexsym}

\usepackage{placeins}

\usepackage{multibib}
\usepackage{natbib}

\raggedright
\newcommand{\IE}{i.e., }
\newcommand{\EG}{e.g., }

\DeclareMathAlphabet{\msfsl}{T1}{cmr}{m}{it}
\DeclareMathAlphabet{\msyf}{OMX}{pcr}{m}{it}

\newcommand{\levelone}[1]{%
\bigskip
\begin{center}
\begin{Large}
\normalfont\scshape #1
\medskip
\end{Large}
\end{center}}

\newcommand{\leveltwo}[1]{%
\bigskip
\begin{center}
\begin{large}
\normalfont\itshape #1
\end{large}
\end{center}}

\newcommand{\levelthree}[1]{%
\vspace{2ex}
\noindent
\textit{#1.}---}

\newcommand{\bibAnnoteFile}[1]{%
\IfFileExists{#1}{\begin{quotation}\noindent\textsc{Key:} #1\\
\textsc{Annotation:}\ \input{#1}\end{quotation}}{}}

\bibpunct[; ]{(}{)}{;}{a}{}{;}

\newcites{sm}{References}

\definecolor{citescol}{RGB}{73,0,165}
\definecolor{urlscol}{RGB}{0,107,124}
\definecolor{linkscol}{RGB}{187,24,0}

\begin{document}

\begin{flushright}
Version dated: \today
\end{flushright}
\bigskip
\noindent RH: GRAPHICAL MODELS IN PHYLOGENETICS

\begin{center}

\noindent{\Large \bf Probabilistic Graphical Model Representation in Phylogenetics}
\medskip

\noindent {\normalsize \sc Sebastian  H\"ohna$^1$, Tracy A. Heath$^2$, Bastien Boussau$^{2,3}$, Michael J. Landis$^2$, Fredrik Ronquist$^4$ and John P. Huelsenbeck$^2$}\\
\medskip

\noindent {\small \it 
$^1$Department of Mathematics, Stockholm University, Stockholm, SE-106 91 Stockholm, Sweden;\\
$^2$Department of Integrative Biology, University of California, Berkeley, CA, 94720, USA;\\
$^3$Bioinformatics and Evolutionary Genomics, Universit\'e de Lyon, Villeurbanne, France;\\
$^4$Department of Bioinformatics and Genetics, Swedish Museum of Natural History, SE-10405 Stockholm, Sweden}\\
\end{center}

\medskip
\noindent{\bf Corresponding author:} Sebastian  H\"ohna, Department of Mathematics, Stockholm University, Stockholm, SE-106 91 Stockholm, Sweden; E-mail: Sebastian.Hoehna@gmail.com.\\

\newpage

\levelthree{Abstract}

Recent years have seen a rapid expansion of the model space explored in statistical phylogenetics, emphasizing the need for new approaches to statistical model representation and software development.
Clear communication and representation of the chosen model is crucial for: (1) reproducibility of an analysis, (2) model development and (3) software design.
Moreover, a unified, clear and understandable framework for model representation lowers the barrier for beginners and non-specialists to grasp complex phylogenetic models, including their assumptions and parameter/variable dependencies. 

Graphical modeling is a unifying framework that has gained in popularity in the statistical literature in recent years. 
The core idea is to break complex models into conditionally independent distributions.
The strength lies in the comprehensibility, flexibility, and adaptability of this formalism, and the large body of computational work based on it. 
Graphical models are well-suited to teach statistical models, to facilitate communication among phylogeneticists and in the development of generic software for simulation and statistical inference. 

Here, we provide an introduction to graphical models for phylogeneticists and extend the standard graphical model representation to the realm of phylogenetics. 
We introduce a new graphical model component, tree plates, to capture the changing structure of the subgraph corresponding to a phylogenetic tree. 
We describe a range of phylogenetic models using the graphical model framework and introduce modules to simplify the representation of standard components in large and complex models.
Phylogenetic model graphs can be readily used in simulation, maximum likelihood inference, and Bayesian inference using, for example, Metropolis-Hastings or Gibbs sampling of the posterior distribution.

\noindent (Keywords: Graphical models, statistical phylogenetics, modularization, tree plate, inference, computation)\\

\newpage


\begin{quotation}
\it
\noindent ...early attempts at reconstructing evolutionary trees 
using computers are leading to a clarification of our 
basic ideas as to how it should be done. It has become 
particularly clear that any attempt at producing an 
evolutionary tree must be based on a specific model, for 
only then can proper statistical procedures be adopted, 
and only then are the assumptions implicit in the 
method clear for all to see.
\begin{flushright}
--- A. W. F. Edwards (1966:440)
\end{flushright}

\rm
\end{quotation}

A basic phylogenetic model consists of a tree with branch lengths and a continuous-time Markov model describing how the characters --- morphological or molecular --- change along the branches of the tree.
Almost every described phylogenetic model fits this theme, which makes it tempting to think that biologists face simple modeling considerations. 
Yet, this is decidedly not the case. 
The variations on the theme of a continuous-time Markov model running along the branches of a tree are seemingly endless. 
From all described models, consider this incomplete list:
TN92,
TN93,
F81,
HKY85,
GTR,
TKF91,
TKF92,
WAG,
BLOSUM,
PAM,
JTT92,
LG08,
REV,
MTREV,
GY94,
MG95,
NY98,
$\mbox{M}_0, \mbox{M}_1, \ldots \mbox{M}_{13}$,
CAT (and CAT again),
MKv,
Dayhoff,
JC69,
K2P, K3P,
ECM,
DEC,
BM,
OU,
EB,
CATBP,
GG98,
TS98,
G01,
UCLN,
UCG,
RLC,
ACLN,
CIR,
and WN.
(The field has inconsistently adopted the practice of naming models with the initials of the authors followed by the year of publication. 
Hence, JC69 refers to the model first described by Jukes and Cantor in 1969.) 
The number of models can be combinatorically increased by the addition of suffixes, such as
 `+I', `+$\Gamma$', `+I+$\Gamma$',  and `+SS', which are different models for accounting for rate variation across characters.
The number of models that are implemented in software and available to the biologist is clearly large. 
Moreover, the scheme adopted by phylogeneticists to name models suggests the field has a considerable degree of opaqueness. 
Clearly, the field could benefit from a generic method ---  a method that can both represent all of the variables contained in a model and their dependencies --- for representing phylogenetic models.

The number and complexity of phylogenetic models presents significant challenges to the biologist.
In some ways, the barriers to understanding a phylogenetic analysis have never been higher.
Software that is intended to simplify phylogenetic analyses can sometimes be counterproductive. 
For example, some software automates the choice of the phylogenetic model for an analysis. 
However, this does not lead to any greater understanding of the assumptions of the analysis by the
user (though such software may ensure a greater overall quality of phylogenetic analysis).
Failure to understand the details of alternative phylogenetic models can lead to innocent
mistakes caused by the confusion of different models having the same name (such as the CAT model, which
is used as a model for rate variation across sites and also as a model for allowing stationary frequencies
to vary across a sequence). 

To address these challenges, we believe the time is mature for the field to adopt a standardized way to describe phylogenetic models. 
Specifically, we suggest following the lead of the statistics literature, where similar problems are encountered, and where graphical models are routinely used to characterize complex 
models \citep{Gilks1994,Lunn2000,Jordan2004,Koller2009,Lunn2009}. 
Graphical models provide a general methodology that works for simple models as well as for large models with thousands, or even millions, of parameters \citep{Jordan2004}. 
Such models are visualized in a simple but comprehensible and exact manner. 
They are independent of the criterion and algorithm used for inference: as long as the model is the same, it does not matter whether inference is performed under the maximum likelihood or Bayesian criterion, or whether Expectation Maximization or Markov chain Monte Carlo sampling is used. 

The paper is divided into three parts, each with a different focus of required expertise in statistical phylogenetics.
We start with a general introduction to graphical models for users of phylogenetic methods.
To this end we model the distribution of the presence/absence of ``the most diverse of bones'', the baculum, in Mammals \citep{Long1968}. 
We draw the corresponding graphical model representation, which we use to introduce the graphical model formalism. 
We progressively transition into phylogenetic models for discrete, continuous and sequence characters but keep the mathematical and technical details to a minimum.

In the second part, we discuss graphical model representations of more typical models used in statistical phylogenetics today.
We introduce the concept of a tree plate, which captures the structure learning part of a phylogenetic model and greatly simplifies the resulting graph.
We also discuss how large and complex phylogenetic model graphs can be modularized to produce more effective views on the overall structure of the model while simultaneously allowing detailed analysis of the model components of particular interest.
In the third part, we present a more formal description of phylogenetic graphical models.
We also provide some well known algorithms on model graphs and relate them to standard algorithms in phylogenetics to demonstrate the benefits of drawing from the vast computational literature on probabilistic graphical models.
We conclude with a discussion on the use and importance of graphical models to the phylogenetics community, and a brief presentation of a software implementation based on phylogenetic graphical models.


\levelone{An Introduction to Probabilistic Graphical Models}

\begin{figure}[!htbp]
\centering
\includegraphics[width=2.5in]{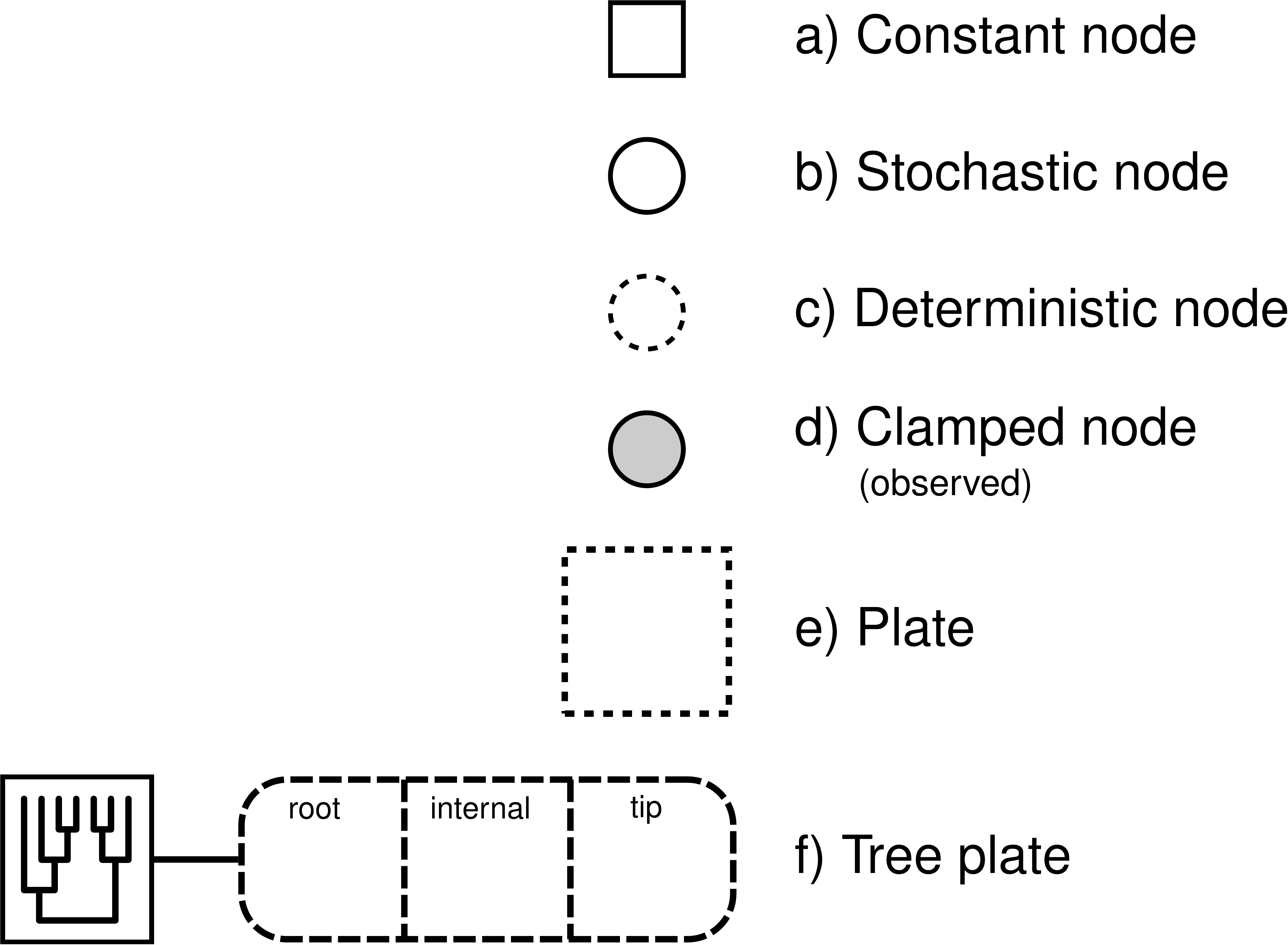} 
\caption{The symbols for a visual representation of a graphical model. 
(a) Solid squares represent \emph{constant} nodes, which specify fixed-valued variables.
(b) \emph{Stochastic} nodes are represented by solid circles. These variables correspond to random variables and may depend on other variables.
(c) \emph{Deterministic} nodes (dotted circles) indicate variables that are determined by a specific function applied to another variable. They can be thought of as variable transformations. 
(d) Observed states are placed in \emph{clamped} stochastic nodes, represented by gray-shaded circles.
(e) Replication over a set of variables is indicated by enclosing the replicated nodes in a \emph{plate} (dashed rectangle). 
(f) We introduce replication over a structured tree topology using a \emph{tree plate}. 
This is represented by the divided, dashed rectangle with rounded corners. 
The subsections of the tree plate demark the different classes of nodes of the tree.
The tree topology orders the nodes in the tree plate and may be a constant node (as in this example) or a stochastic node (if the topology node is a solid circle).}
\label{fig:GM_notation}
\end{figure}

The graphical model framework provides a valuable set of tools for visually representing models. 
The various components of a graphical model representation are defined in Figure~\ref{fig:GM_notation}. 
The following examples will introduce each of the elements needed for constructing model graphs. 

\medskip

\FloatBarrier
\leveltwo{A non-phylogenetic presence/absence model}

The \textit{os penis} of mammals, or \emph{baculum}, has an uneven taxonomic distribution. It occurs in five orders of mammals \citep[Carnivora, Chiroptera, Insectivora, Primates, and Rodentia;][]{Patterson1982} but is absent
in all other mammalian orders, including marsupials and monotremes.
The evolution of this character has been studied to determine potential use of the presence of the baculum.
Potential hypotheses for the evolution of the baculum include (1) a purpose as a stiffener for species with extended  intromission, (2) to assist in sperm transport, or (3) to provide rigidity to stimulate
female ovulation \citep{lariviere02}.
Here we consider some of the modeling considerations for a phylogenetic analysis of this character.
We  choose to use Bayesian methods to conduct these inferences, which means that we will need to specify prior probability distributions for the variables of our models.
To simplify our analyses, we will sample five species: a dog, a bat, a rat, a human, and a koala. 
The supplementary material presents similar analyses with a much better taxonomic sampling of 274 species.

\begin{figure}[!htbp]
\centering
\includegraphics[width=5in]{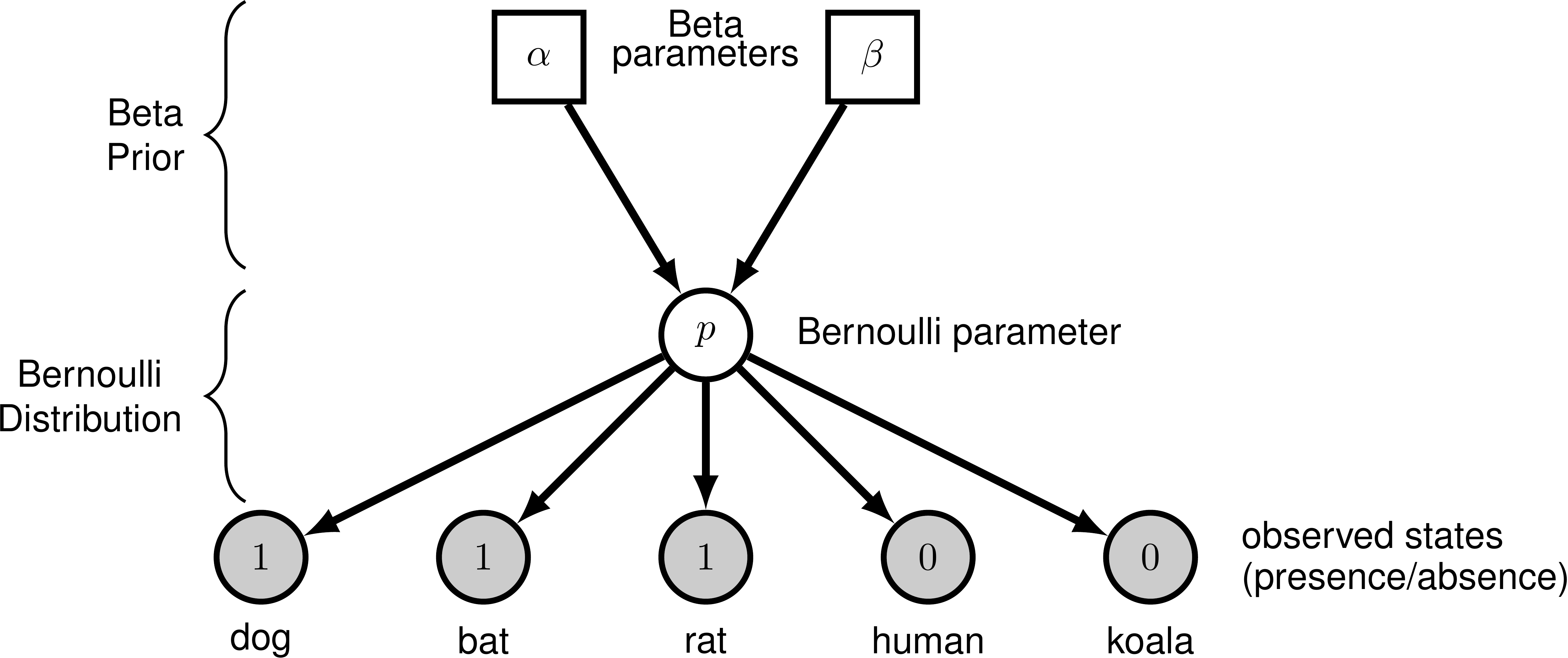} 
\caption{An explicit graphical model of the distribution of a binary trait. 
Descriptions of the objects have been added for pedagogical purpose.
The presence or absence of the binary trait is assumed to follow a Bernoulli distribution with parameter $p$. 
This parameter is equal to the probability of the presence of the baculum in an independently sampled species.
We place a Beta prior density on the Bernoulli distribution parameter, such that $p \sim \mbox{Beta}(\alpha,\beta)$, where $\alpha=1$ and $\beta=1$ are the shape parameters of the Beta distribution.
This probability density is defined on the interval $[0,1]$, thus $0 \leq p \leq 1$. 
}
\label{fig:binaryFig1}
\end{figure}

Our first attempt at modeling the distribution of the baculum assumes that all species are independent of one another but share the same probability of having a baculum. 
The probability of obtaining a baculum follows a Bernoulli distribution:
with some probability $p$, a species receives a baculum, and with probability $1-p$, it does not. 
We specify a Beta prior probability distribution on the value of $p$, which is adequate for values between 0 and 1. 
This Beta distribution itself has two parameters $\alpha$ and $\beta$, both of which we set to 1. 
This choice conveys our lack of knowledge regarding the value of $p$ as it generates a uniform distribution over $[0,1]$.
The corresponding model is represented in Figure~\ref{fig:binaryFig1}. 
The graph is composed of nodes and arrows joining them.
The nodes correspond to variables of our model, such as the Bernoulli parameter $p$, and the arrows correspond to dependencies between the variables.
For instance, the top arrows show that the value of $p$ depends upon the parameters $\alpha$ and $\beta$.
In fact, the dependency structure in a graphical model is easily read by following the arrows backwards from a dependent variable, to which an arrow points, to the variable it depends upon.
In contrast, if we were to simulate data according to a graphical model, the flow of the simulation would be forwards along the direction of the arrows.

In Figure~\ref{fig:binaryFig1}, we have chosen a somewhat verbose description, with labels next to the nodes and arrows.
In more complex models, it is customary to dispense with these names and only rely on the symbols inside the nodes to avoid cluttering.
In the same manner that algebraic symbols are indispensable for solving complex equations, the use of short symbols is indispensable for representing complex probabilistic models.
However, the representation of nodes in the graph carries additional information: square nodes are constant nodes (\EG $\alpha$ and $\beta$) that depend on no other node (thus sometimes called source nodes), and circular nodes are not constant (see Fig.~\ref{fig:GM_notation}). 
In the present model, all circular nodes are stochastic, \IE each circular node corresponds to a random variable, whose value comes from a probability distribution. 
Some of our stochastic nodes have been shaded (Fig.~\ref{fig:binaryFig1}), which means that they have been ``clamped". 
A clamped node is a stochastic node whose value has been observed and thus data are attached to the node. 
In our case, the bottom nodes have been clamped because they correspond to anatomical observations in the species of interest.
We inferred the value of $p$ in this model using a Markov chain Monte Carlo (MCMC) algorithm, and found that its value was $0.57$ with 95\% highest posterior density (HPD) interval of $[0.23,0.88]$ 
($\hat{p} = 0.48 \text{, HPD}=[0.42,0.54]$ on the larger data set, see Sup. Mat.).

\leveltwo{A simple phylogenetic model}

\begin{figure}[!htbp]
\centering
\includegraphics[width=\textwidth]{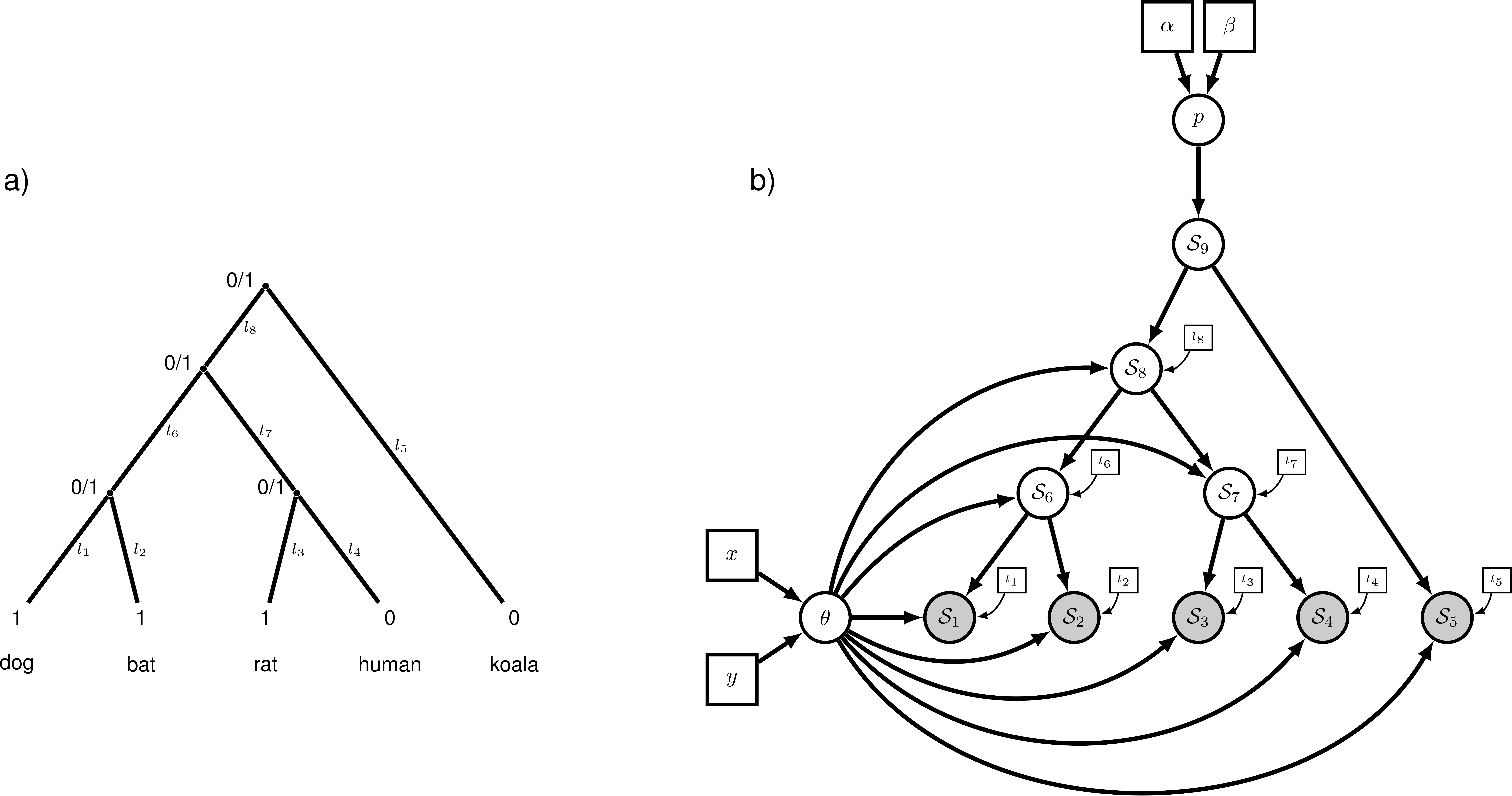} 
\caption{The evolution of a single binary character represented as a phylogenetic graphical model. 
a) The phylogenetic relationships of the five mammalian species. 
The observed state of the character (1: presence or 0: absence of the baculum) is given for each species. 
Other states at the internal nodes represent the unknown ancestral state.
The branches of the tree ($1,\ldots,8$) are labeled and assigned a fixed length ($l_1,\ldots,l_8$). 
b) The corresponding graphical model, in which the species tree topology is still evident.
We represent the state for each node with generic notation: $\mathcal{S}_1$ is the presence/absence state for node $1$. 
The clamped nodes, in grey, indicate observed states, whereas unobserved states for ancestral species are in white. 
Constant nodes indicate fixed/known branch lengths. 
Under this model, the state for the root of the tree ($\mathcal{S}_9$) is drawn from a Bernoulli distribution with probability $p$. 
A Beta prior is assigned to the parameter of the Bernoulli distribution so that $p \sim \mbox{Beta}(\alpha,\beta)$, where the parameters of the Beta distribution are constant nodes and assigned fixed values. 
The states of the nodes descended from the root of the tree ($\mathcal{S}_1,\ldots,\mathcal{S}_8$) are dependent on the equilibrium frequency parameter ($\theta$) and their respective branch lengths (constant nodes $l_1,\ldots,l_8$). 
A second Beta distribution is applied as a prior on the parameter $\theta$, where $\theta \sim \mbox{Beta}(x,y)$. 
}
\label{fig:EvolBaculum}
\end{figure}

Obviously this model fails to take into account the known phylogenetic structure underlying the distribution of the baculum among Mammalian species.
We therefore propose a second model, in which the presence/absence of a baculum is represented as a binary character evolving along the Mammalian phylogeny.
The evolution of this binary character is modeled by a continuous time Markov process, which only needs two parameters, the equilibrium frequency $\theta$ of character ``1'' and the set of branch lengths, assuming that the Markov process is parametrized in units of time (no transformation of the branch lengths is necessary).
At the root of the tree, we need to specify a prior probability distribution over the parameter $p$ representing the probability of the presence/absence of a baculum.
As for the first model, we use a Beta prior distribution with parameters $\alpha$ and $\beta$ both set to 1 ($p \sim \mbox{Beta}(\alpha,\beta)$).
We also need another parameter $\theta$ for the equilibrium frequency of the state 1, parameterized the same way as $p$.
We use the dated phylogeny of \cite{Reis2012}, pruned to contain only the five species of interest or the 274 species in our dataset (supplementary material).
We assume this phylogeny is known without error (Fig.~\ref{fig:EvolBaculum}a).
Comparing Figure~\ref{fig:EvolBaculum}a and Figure~\ref{fig:EvolBaculum}b demonstrates how the structure of the phylogenetic tree (partially) forms the structure of the graphical model.
The structure of the phylogenetic tree can be recovered as a central subset of the graphical model, because each node of the phylogenetic tree is a stochastic variable in our model, taking values 0/1, and depending only on its parent node, the branch length and on the parameter $\theta$ of the continuous time Markov process.
In the supplementary material we provide scripts for performing Bayesian inference with this model and the complete data set, and we are proud to report that there is a 50/50 chance that the ancestor of Mammals had a baculum.

We believe such a graphical model representation is a very powerful pedagogical construct, as it displays the entire structure of our probabilistic model.
It makes it easy for a student or a reviewer to identify key assumptions made by this model.
For example, although the evolutionary process is the same along all branches of the phylogenetic tree, the model is not stationary, because the root has an extra parameter for the probability of presence/absence of the baculum ($p \neq \theta$).
However, even though our model is simple and contains few species, our graph is already quite busy.
Clearly, an explicit representation is impractical for large numbers of characters, or for much more complex models, and some factorization needs to be performed.

\FloatBarrier
\leveltwo{Using plates to represent repetition in the graph}

\begin{figure}[!htbp]
\centering
\includegraphics[width=4in]{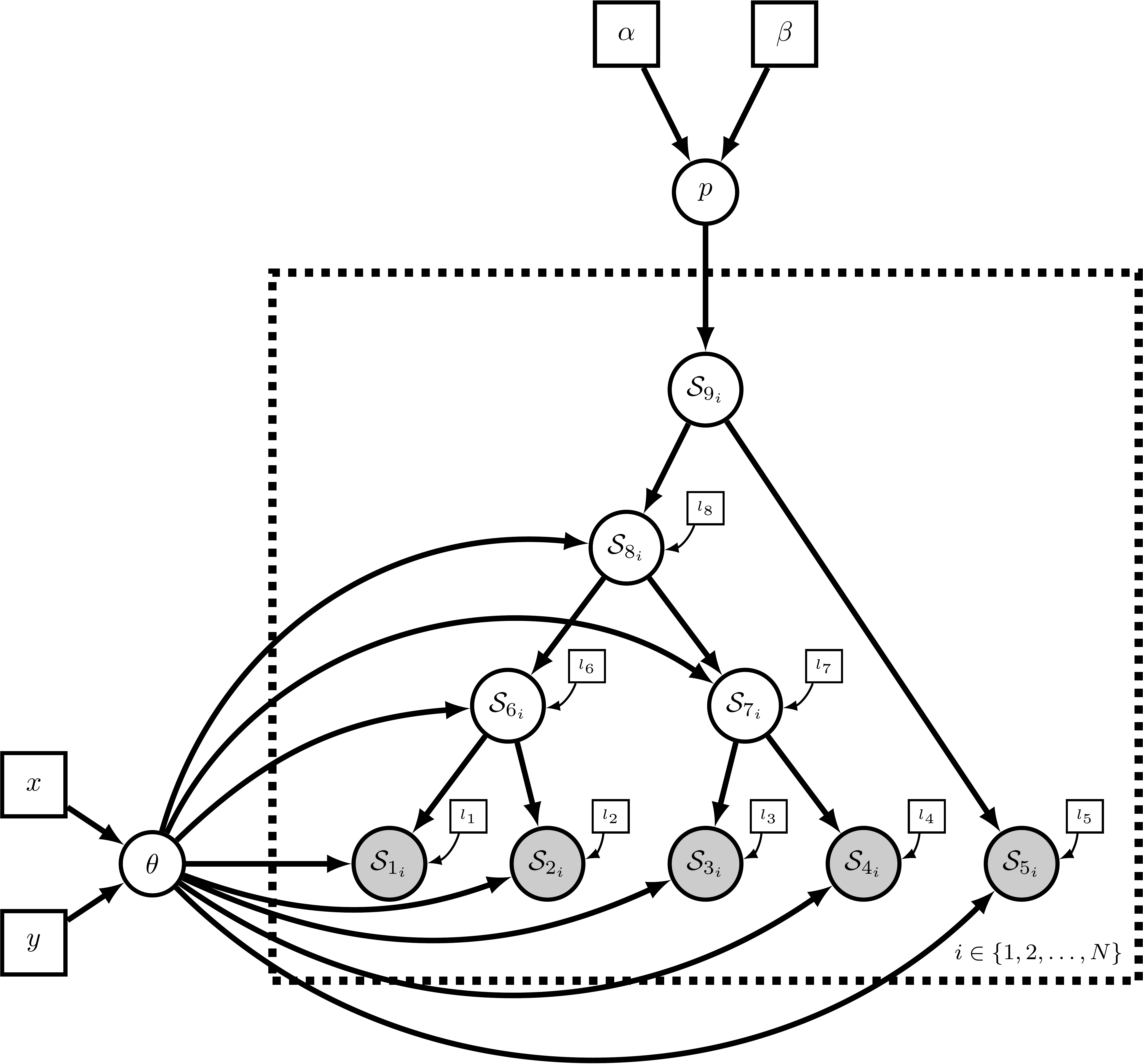} 
\caption{A phylogenetic graphical model of $N$ independently evolving binary characters.
When sampling $N$ different binary characters for each extant species, we assume that these characters are independent and identically distributed. 
Thus the model for each character is the same as in Figure~\ref{fig:EvolBaculum}b. 
Yet, the state for each character $1,\ldots,N$ can be different. 
We use the \textit{plate} notation to represent repetition over a vector of elements. 
In this figure, the dashed box and the iterator $i$ indicate the replicated variables. 
Thus, the plate represents separate variables of binary character evolution for $i$ in characters $1,2,3,\ldots,N$.
} 
\label{fig:EvolBaculumPlate}
\end{figure}

Data are inherently repetitive and this feature must be efficiently captured by a graphical model.
What if, in addition to the baculum, we also wanted to analyze the distribution of the \textit{os baubellum}, found in females, and of a few other binary characters? 
Our model graph would quickly become cluttered. 
To circumvent this problem, the graphical model literature uses \textit{plates} to represent iteration \citep{Jordan2004,Koller2009}. 
Plates are represented as a dotted rectangles on top of which repetitive nodes are placed (Fig.~\ref{fig:EvolBaculumPlate}). 
In a corner of the plate, the number of repetitions --- in our case binary characters --- is given. 
Assuming we analyze $N$ binary characters using the same underlying Markov process running along the branches of the phylogenetic tree, we need to put the entire phylogenetic tree on the plate.
In fact, the variables of both the leaves and the internal nodes of the phylogenetic tree differ for each character in our data matrix, because they correspond to different characters and their ancestral states, though the ancestor/descendant relationships remain unchanged.
We chose to leave the parameters of the probability of presence/absence at the root off the plate, which means that we assume the probability of presence at the root is the same for all $N$ characters. 
Similarly, we have left $\theta$ off the plate, assuming that all $N$ characters evolve under the same transition probabilities. 
These very strong and debatable assumptions are highlighted by the graphical model representation.

\FloatBarrier

Graphical models are high-level representations that do not depend on details of the model, such as which distribution is applied to a variable. 
As a result, similar models will have similar structural representations. 
We provide in the supplementary material the example of a Brownian motion model of the evolution of continuous characters to convey this point \citep{Felsenstein1985}, and show here in more detail the example of a model of sequence evolution. 

\FloatBarrier
\leveltwo{A general-time-reversible model for sequence evolution}

\begin{figure}[!htbp]
\centering
\includegraphics[width=\textwidth]{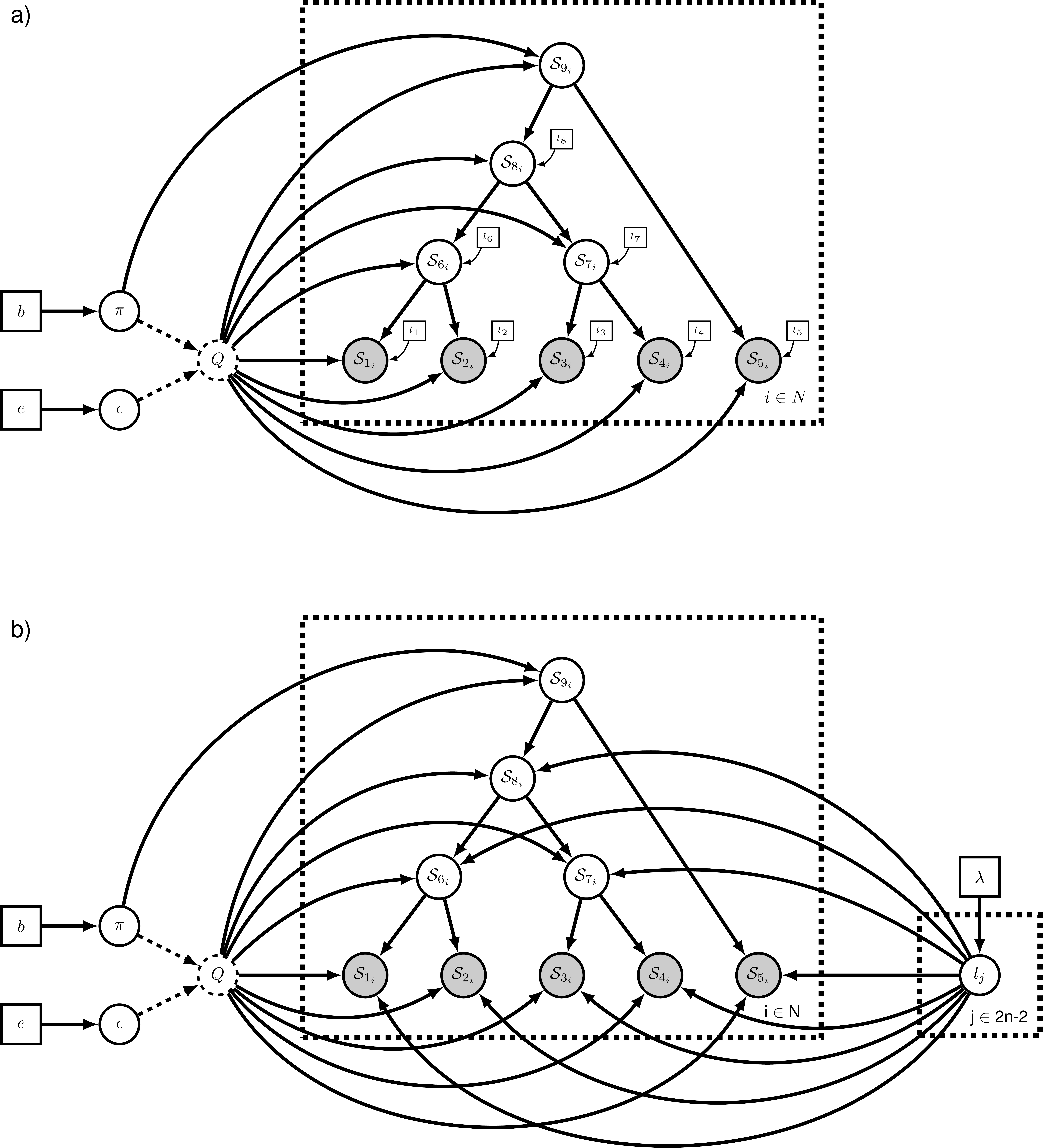} 
\caption{
Explicit graphical model representation of a GTR model with a fixed tree topology.
For pure convenience we show here rooted trees that demonstrate the similarity to previous figures.
The model of character evolution is a continuous time Markov model parameterized by an instantaneous rate matrix.
The rate matrix $Q$ is a deterministic variable computed by multiplying the base frequencies $\pi$ with the exchangeability rates $\epsilon$.
A Dirichlet distribution is applied as the prior distribution on both the base frequencies $\pi$ and the exchangeability rates $\epsilon$.
a) A GTR model with fixed branch lengths.
b) A GTR model with estimated branch lengths.
Each branch length is independent and identically distributed under an exponential distribution.
}
\label{fig:gtrExplicitBlens}
\end{figure}

One of the most popular models of sequence evolution is the general time reversible (GTR) substitution model \citep{Tavare1986}.
Here we give a simple example of a GTR model for a fixed, non-clock tree with fixed branch lengths. 
In this case branch lengths are not defined in units of time as in the previous examples, but instead in expected numbers of substitutions and for simplicity we consider that we have some trustworthy exterior information about them.
The resulting graphical model is depicted in Figure~\ref{fig:gtrExplicitBlens}a, and is very similar to the previous figures for the binary and continuous characters (Fig.~\ref{fig:EvolBaculumPlate} and Fig.~S.1). 
The tree sits on a plate because it is replicated for $N$ sites. 
In this example, every character evolves under a continuous time Markov model with transition rate matrix $Q$ and branch length $l_{j}$ where $j$ denotes the index of the branch. 
The transition rate matrix $Q$ is defined as a deterministic function computing the transition rates by multiplying the exchangeability rates with the base frequencies. 
This deterministic computation is represented differently from other dependencies among nodes, with a dashed arrow pointing into a dashed node.
The visually distinctive representation of deterministic nodes is used to show that the value of a variable is deterministically computed from the values of parameters it depends upon, \IE by a transformation of the parameters.
This completes our compendium of nodes used in graphical models.

Of course, branch lengths are often estimated instead of being considered constant. 
In a Bayesian context one would then have to provide priors for the branch lengths.
In Figure~\ref{fig:gtrExplicitBlens}b we show the graphical model corresponding to a GTR model for a fixed tree, but in which branch lengths are estimated. 
As is customary, we use an exponential prior on branch lengths, with parameter $\lambda$. 
Although this model is more complex and its representation busier, the structure of the tree can still be recovered from the graph. 
In fact, all phylogenetic examples provided thus far share a common structure, namely the underlying tree structure. 
We believe these strong similarities contribute to making graphical model representation powerful for teaching and understanding phylogenetic models. 
With some use, it becomes easy to identify the unique parts of a particular model, and the parts that relate it to alternative models.

To summarize this introduction to the graphical model framework (see Fig.~\ref{fig:GM_notation}), we have constant variables represented with square nodes, and variables whose value can change --- during simulation or inference --- represented with circular nodes. 
Circular nodes can be stochastic, with solid lines, or deterministic, with dashed lines. 
Arrows pointing into variable nodes represent conditional dependencies, visualized in solid or dashed lines depending on the nodes they point into. 
In addition, we have plates, that convey the concept of repetition. 
In more formal terms, nodes placed on a plate correspond to independent, identically distributed variables.
This list of graphical representations is commonly used in the statistics literature \citep{Lauritzen1996,Jordan2004,Koller2009}. 
For phylogenetics, where we often handle phylogenetic trees that can contain large numbers of nodes, and whose topology is often unknown and needs to be estimated, other constructs are needed.
Those constructs, \EG \textit{tree plates}, are introduced in the next section.

\FloatBarrier
\levelone{Phylogenetic Model Graphs}

Ordinary graphical models are impractical for describing realistic phylogenetic models for two reasons:
First, visual representations of these models become crowded as the number of tips grow, and essential information may become buried in a litany of details.
Second, ordinary graphical models fail to represent topological (structural) uncertainty because the dependency structure of the nodes in the graph, corresponding to the phylogenetic tree topology, is fixed.
We solve both problems by adding a new element to the list of graphical model conventions: a \emph{tree plate}.

\leveltwo{Tree Plates}

\begin{figure}[!htbp]
\centering
\includegraphics[width=\textwidth]{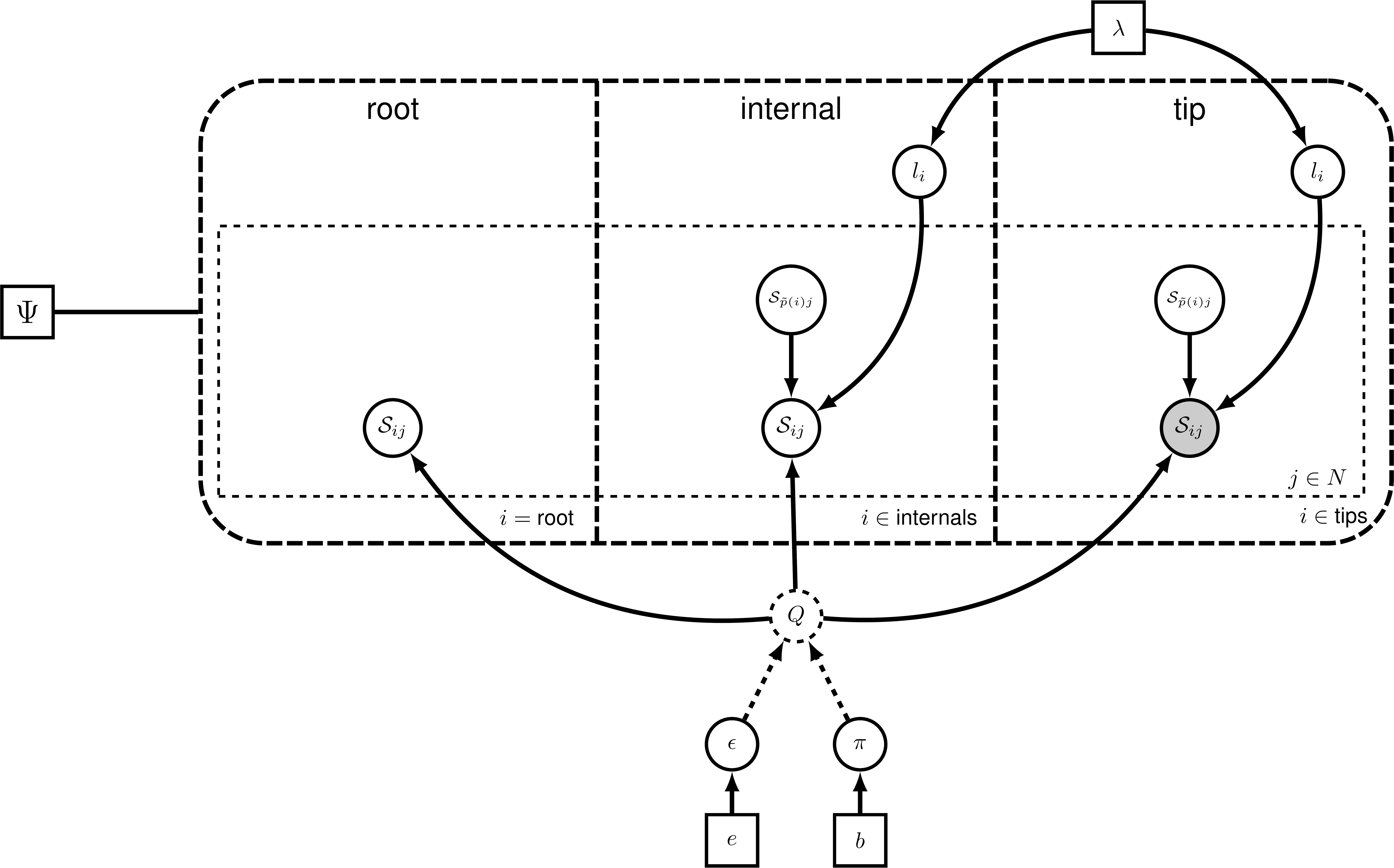}
\caption{
Simplified representation of the GTR model of Figure~\ref{fig:gtrExplicitBlens}b using a tree plate.
The tree plate, a big dashed box, divides the nodes into three classes: the root node, internal nodes and tip nodes.
The character state variables are named $S_{ij}$ where $i$ denotes the $i^{\text{th}}$ node and $j$ the $j^{\text{th}}$ site. 
The root node does not have a parent node in the tree while the other nodes do.
The internal nodes and the tip nodes depend on the ancestral states.
The ancestral variable of node $i$ is obtained using the parent indicator function $\tilde{p}(i)$.
Tip nodes are clamped and thus shaded.
A tree topology is attached to the tree plate via the tree variable shown on the left.
The tree variable informs the plate of the structure and if the tree variable changes, the structure of the resulting graph changes too.
}
\label{fig:gtrgTreeplate}
\end{figure}

A tree plate is very similar to a plate.
However, where a plate symbolizes repetition of a particular element in the model, a tree plate symbolizes recursion: a given variable depends upon a conceptually similar variable.
Recursion is a concept that fits naturally within a tree, given that many nodes in a tree are the child of a parent as well as the parent of a child.
Naturally, recursive constructs need initiating and terminating: the recursive description of a tree starts at the root node, and terminates at the tips. 
This suggests that a tree plate needs to account for three classes of nodes at least: the root node, internal nodes, and tip nodes.
Contrary to internal nodes and tips, the root node does not depend on a parent node in the tree. 
Contrary to internal nodes, tips are often clamped to observed values. 
Figure~\ref{fig:gtrgTreeplate} represents a tree plate as a big, rounded rectangle divided into three parts, one for each class of nodes, with a tree variable attached to it providing the structural information. 
Parent-child relationships are handled by special functions for the indexing of the parent node: $\tilde{p}(j)$ represents the parent in the tree of node $j$ and $\tilde{c}(i,j)$ represents the $i^{\text{th}}$ child of node $j$ in the tree.
A comparison between Figure~\ref{fig:gtrExplicitBlens} and Figure~\ref{fig:gtrgTreeplate} shows how a tree plate can simplify the representation of a phylogenetic model, and how it interacts nicely with a plate.
The example is extended in the Suppl. Mat. by the commonly used mixture model for rate variation across sites, the GTR+$\Gamma$ model \citep{Yang1994,Yang1996} (see Figure~S.2).

Importantly, the recursive representation of a tree plate protects it against cluttering as the size of the tree grows: no matter how many tips are included in the tree, these three classes of nodes are enough to describe most phylogenetic models.
More classes are only needed when the model further distinguishes between nodes, for example when different models of sequence evolution are associated with different subtrees or when particular nodes are associated with time calibration information.
The tree plate also adequately addresses the representation of topological uncertainty.
Because the tree plate uses a high-level, recursive representation of a tree, it transcends a specific tree topology and instead allows any tree topology. 
Only the specific value of the tree variable ordering the tree plate reveals the actual graphical structure of the model.

\leveltwo{Modularization}

Although tree plates simplify the representation of a phylogenetic model, visualization remains a challenge for the most complex models.
As an example consider the common case of a multi-locus analysis using a mutli-species coalescent model, an uncorrelated relaxed clock and a GTR+$\Gamma$ substitution model, which would be represented by Figure~S3, Figure~S4 and Figure~S2 merged together.
Clearly, such a figure would be overwhelming.
Ideally, one would like a method that allows one to quickly convey the bigger picture while allowing parts of special interest to be exposed in all the necessary detail.
For instance, it is common practice to create new phylogenetic models by combining existing model components, possibly in new patterns, with new components.
In such situations, it is practical for a computational phylogeneticist to use a simplified, high-level representation of the complete model graph, and focus on the model subgraph(s) of interest in the discussion of the novelties.
Similarly, an evolutionary biologist might be interested in effectively communicating the crucial differences in the overall structure of some models without going into all the model details.
To address these challenges, we propose a factorization of phylogenetic model graphs into modules, each of which corresponds to a subgraph of potential interest.
The modules can be collapsed into simple graphical objects to allow compact high-level representation of a large model.
One or more modules can also be expanded to expose all the details of the corresponding model subgraph.
This allows one to communicate both the overall structure of a large model and the details of the model components of particular interest.

\levelthree{Module decomposition of a phylogenetic model}

When discussing a set of related complex models, it is sufficient to describe the detailed structure of common modules once.
After that, new model variants can be characterized by simply specifying the subgraph structure of the modules that differ.
An overview of the structure of a large and complex model is obtained by breaking it into appropriate modules and representing all modules in their collapsed form.
This assumes that there is some common understanding of the subgraph structure within each of the modules, for instance by reference to previous papers on these modules.

\begin{figure}[!htbp]
\centering
\includegraphics[width=\textwidth]{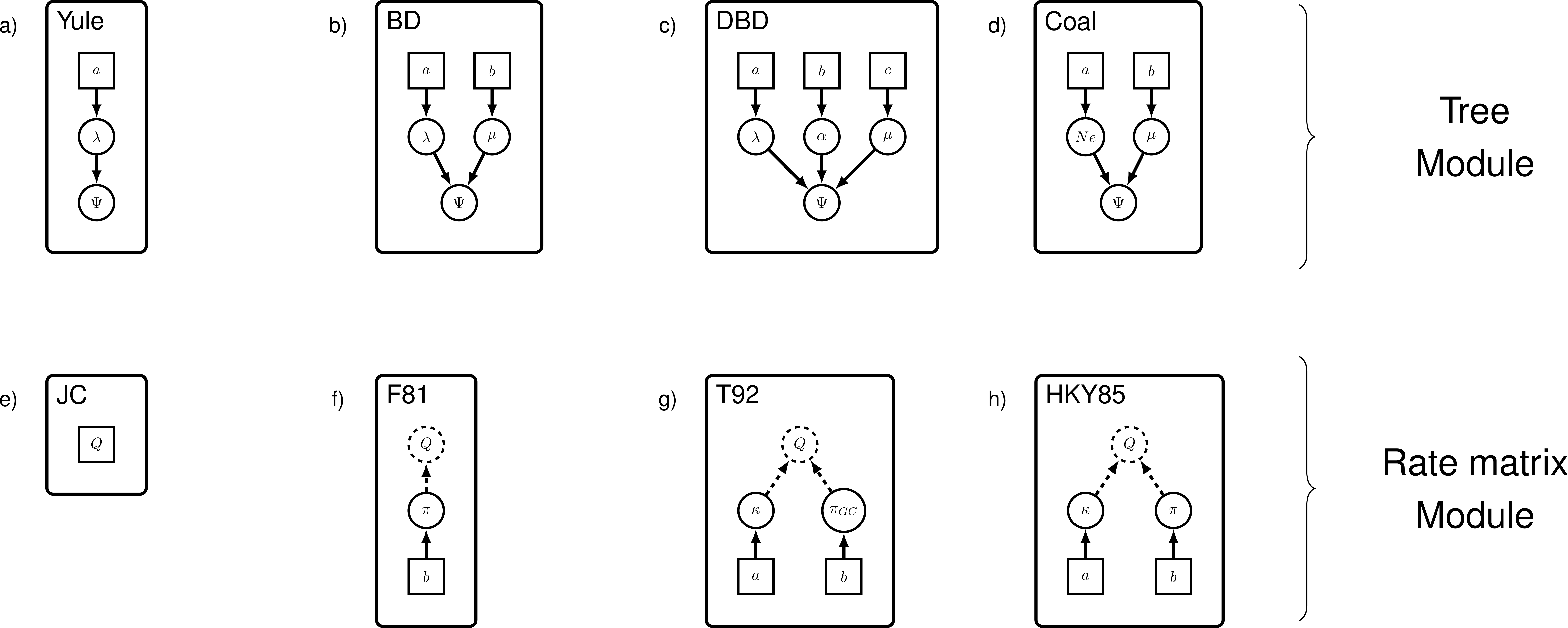} 
\caption{
Top panel:
Module representation of different tree priors with $\Psi$ as a pivot node: 
a) Yule process \citep{Yule1925}, 
b) constant rate birth-death process \citep{Nee1994b}, 
c) decreasing speciation rate birth-death process (speciation rate: $\lambda * \exp(-\alpha t)$)  
and d) Coalescent process \citep{Kingman1982}.
Bottom panel:
Different rate matrix modules with $Q$ as a pivot node:
e) Jukes-Cantor rate matrix where all exchangeability rates and all base frequencies are equal \citep{Jukes1969},
f) F81 rate matrix where all exchangeability rates are equal but the base frequencies are drawn from a Dirichlet distribution \citep{Felsenstein1981},
g) T92 rate matrix with a parameter for the frequency of the GC content $\pi_{GC}$ and a transition-transversion rate \citep{Tamura1992}
and h) HKY85 rate matrix with the base frequencies drawn from a Dirichlet distribution and an estimated transition-transversion rate \citep{Hasegawa1985}.
}
\label{fig:module_alternatives}
\end{figure}

\begin{figure}[!htbp]
\centering
\includegraphics[width=0.9\textwidth]{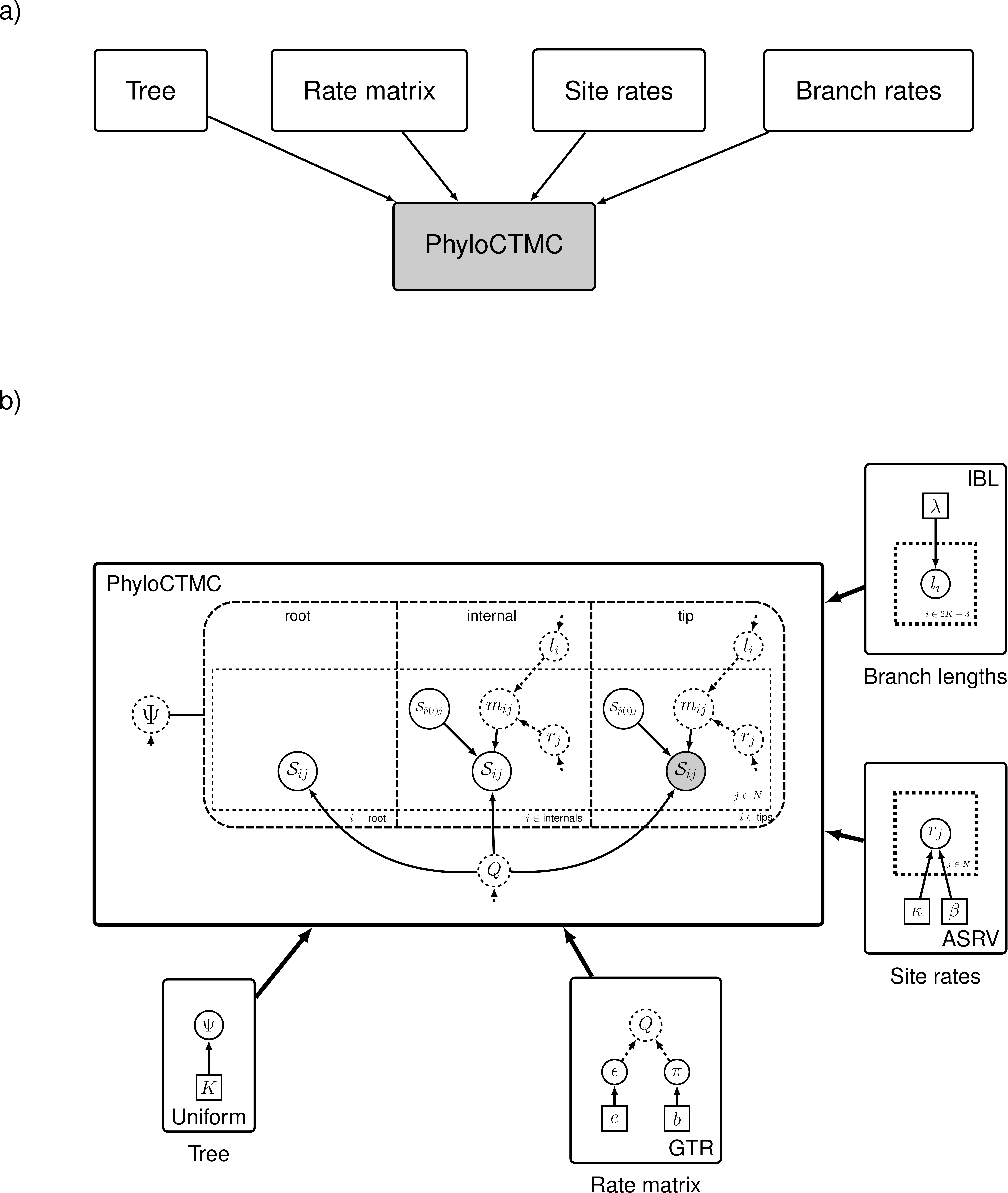} 
\caption{
The graphical model of Figure~\ref{fig:gtrgTreeplate}, a GTR+$\Gamma$ model, represented in modular form.
a) The model is broken into five different modules: Tree, Rate matrix, Site rates, Branch rates and PhyloCTMC (Continuous Time Markov Chain).
By representing all modules in collapsed form, we obtain a compact high-level visualization of the model.
Arrows point from upstream to downstream components in the complete model graph.
b) By expanding the modules to expose the model subgraphs they contain, we obtain a detailed description of the model.
Note that the four upstream modules (Tree, Rate matrix, Site rates, and Branch rates) are all named after the corresponding pivot variable.
Also note that the symbols used for pivot variables are matched across connected modules, both by name and by plate or tree plate indices.
Tiny arrows aid the search for pivot variables.
}
\label{fig:modules}
\end{figure}

To factorize a complex phylogenetic model graph into modules, we introduce a new element to our graphical models: pivot nodes.
Pivot nodes allow a single random variable to appear simultaneously in several subgraphs. 
This is useful for explicitly defining how modules interface.
The pivot nodes may represent unique (single) variables (\EG the tree variable or the rate matrix variable) or they may represent a collection of variables (\EG the set of branch rates).
Suitable pivot nodes are variables that differ in alternative models (see Fig.~\ref{fig:module_alternatives}).
After a pivot has been identified, the model graph is partitioned into two modules, one module representing the upstream structure of the model graph and the other module representing the downstream structure, with the pivot node being represented in both modules (see Fig~\ref{fig:modules}).

As a high-level representation of a module, we use a solid rectangle containing appropriate text describing the module.
An upstream module is connected to a downstream module by an arrow pointing to the latter and thus depicts the dependency structure.
When a module is expanded to expose the details of the model subgraph it contains, we use the standard phylogenetic graphical model conventions.
The connections between the modules are made explicit by using the same variable names and plate indices in all modules.
Across alternative complex models, a pivot node may be stochastic, deterministic or constant (see  Fig.~\ref{fig:module_alternatives}e-h).
To preserve the subgraph representation of the downstream module in such cases, we suggest using a deterministic node representation of the pivots in the downstream module.
This is compatible with the graphical model conventions, in that the value of the pivot variable in the downstream module can always be obtained as an identity transformation of the corresponding variable in the upstream module, regardless of whether the pivot variable is constant, deterministic or stochastic in the latter.
Moreover, a pivot variable in a downstream module may be used as a collection of variables (\EG the branch rates) but the upstream module only specifies a unique variables (\EG an overall clock-rate).
In this case the same variable is used for each index as done by a deterministic replication of the variable.

The most practical choice of pivot variables and the corresponding modularization of phylogenetic model graphs is not obvious in all cases.
These problems will undoubtedly be discussed in the phylogenetics community, and we expect that the use of modules will evolve to some extent over time.
However, we propose some obvious pivot variables and associated modules here, as a starting point for further discussion.
Figure~\ref{fig:modules} presents one potential module factorization of the GTR model (Fig.~\ref{fig:gtrgTreeplate}).

\levelthree{PhyloCTMC module}
The \emph{PhyloCTMC module} is commonly the core of a phylogenetic analysis.
Typically, the nodes representing the leaves of a phylogenetic tree would be clamped to the observations contained in a character matrix, such as a set of aligned DNA sequences.
A standard phylogenetic model contains a single PhyloCTMC module, but more complex models might have replicated PhyloCTMC modules, \EG one PhyloCTMC module for each gene using different rate matrices.
It may be used in a simple model where all characters evolve homogeneously or it may be extended by, \EG using site-specific rate-multipliers \citep{Yang1994,Yang1996}, branch-specific rate-multipliers \citep{Thorne1998}, branch-specific substitution rate matrices \citep{Yang1995,Galtier1998} and site-specific tree topologies \citep{Boussau2009}.
Some of these extension are described in the next modules.

\levelthree{Tree module}
The \emph{tree module} represents the subgraph describing the tree model, \IE the model of tree topology and possibly also associated branch lengths or node ages.
A tree module could be used to represent a fixed topology with or without fixed branch lengths.
More commonly, the tree module would be used to specify a prior distribution on trees or topologies.
In the main example shown here (Fig.~\ref{fig:modules}), the tree module is a uniform distribution on unrooted topologies. 
Alternative tree modules (Fig.~\ref{fig:module_alternatives}a-d) include the Yule or pure-birth process \citep{Yule1925}, the birth-death process with a constant speciation and extinction rate \citep{Thompson1975,Nee1994b}, the decreasing speciation rate birth-death model (SPVAR in \cite{Rabosky2008}), and the coalescent process \citep{Kingman1982}.

\levelthree{Branch lengths or branch rates module}
Two other suitable pivot variables are the branch length and branch rate variables, producing an upstream \emph{branch lengths module} and \emph{branch rates module}, respectively.
Note that a branch lengths module and a branch rates module are both simple scaling factors of the branch lengths and thus can be applied interchangeably.
For instance, we might consider all branch lengths drawn from a common distribution, as in our example model (Fig.~\ref{fig:modules}), or we might consider a more complex model where branch lengths are drawn from separate distributions sharing a parameter drawn from a common distribution.
A branch rates module would specify the model on a rate multiplier applied to the branch lengths.
The multiplier could either apply to all branches in the tree (if it were represented by a single variable) or applied per branch (if it were replicated across the tree plate).
The branch rates module would be a central component of relaxed clock models.
It could also be used to describe a rate multiplier for different gene partitions, in which case the pivot variable would be replicated across the gene partition plate in the downstream core of the phylogenetic model, rather than across the tree plate as in a relaxed clock model.
An example of the branch rates module for the autocorrelated lognormal distributed rates \citep{Thorne1998,Heath2012a} is given in the supplementary material (see Suppl.~Mat. Fig.~S.3).

\levelthree{Rate matrix module}
The instantaneous rate matrix of the substitution model is the pivot variable of the \emph{rate matrix module}.
The pivot variable may be unique and apply to all sites and branches in the PhyloCTMC module in a branch-homogeneous substitution process.
It may also be replicated in the tree plate, \EG across branches, in which case each branch would potentially be characterized by a unique substitution process \citep{Yang1995,Galtier1998,Groussin2013}, across sites \citep{Lartillot2004}, or according to models with explicit dependencies between neighboring branches \citep{Blanquart2006}.
In all such cases, the rate matrix module would describe the dependency structure of the rate matrix variable.
For instance, a GTR rate matrix would be computed deterministically from a vector of stationary state frequencies and a vector of exchangeability rates (Fig.~\ref{fig:modules}b).
A large portion of the phylogenetic model space considered currently can be characterized by variations on the subgraph structure corresponding to the rate matrix module.
Some examples are shown in Figure~\ref{fig:module_alternatives}e-h.

\levelthree{Site rates module}
The final pivot variable we consider here is the variable used to model heterogeneity of rates across sites \citep{Yang1994,Yang1996} embedded in the \emph{site rates module}.
Commonly the rates for each site are considered drawn from a gamma distribution.
The distribution is typically discretized for computational reasons.
Interestingly, a discrete gamma model could be explicitly described by assuming that the site rates are drawn from a discrete mixture of rates, each rate being deterministically derived by computing the appropriate discrete representation of a gamma distribution (see Suppl.~Mat.~Fig.~S.2).
Alternatively, each site rate may be drawn directly from the gamma distribution, as in our example (Fig.~\ref{fig:modules}b).
Other distributions than the gamma can be used for the rate variation across sites, sometimes leading to better results \citep{Mayrose2005a}. 
In addition to models based on simple continuous distributions, any mixture model of rates \citep{Pagel2004,Pagel2005} would be eligible for the site rates module.
For instance, a standard model considered in the literature is a mixture of invariable sites (rate zero) and gamma-distributed rates.

\levelthree{High-level modular graphs}

\begin{figure}[!htbp]
\centering
\includegraphics[width=\textwidth]{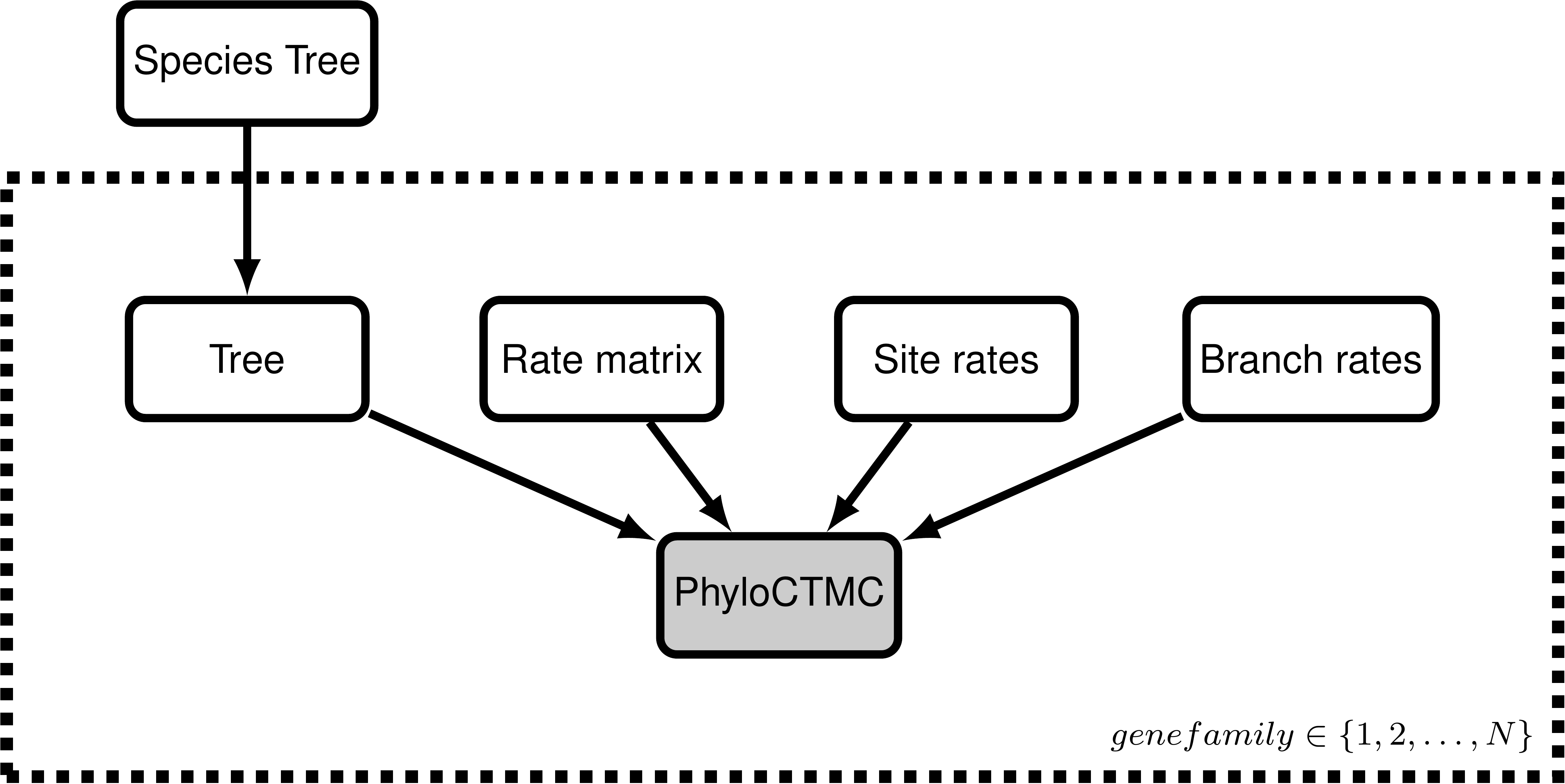} 
\caption{
Module representation for a species tree-gene tree model.
We simply extend the previous phylogenetic model by substituting the simple tree module by a modular representation of a species tree prior and a gene tree distribution given the species tree.
The gene tree with the entire substitution process sits on a plate representing that the model is repeated across genes.
The PhyloCTMC module is shaded to reflect the fact that it is clamped to observations.
}
\label{fig:spTreeGeneTreemodule}
\end{figure}

We end this section by a simple example illustrating the power of high-level modular graphs in summarizing the essential structure of a large and complex model.
For this example, let us consider a model where we want to simultaneously estimate a set of gene trees and the species tree into which they fold.
The high-level representation is obtained by extending the previous module graph with a species tree --gene tree model (Fig.~\ref{fig:spTreeGeneTreemodule}).
The gene-tree part of the model sits on a plate representing the replication over genes.
The PhyloCTMC module is shaded to reflect the fact that it is clamped to the observations, \IE the sequences at the leaves.
All gene trees depend on a single species tree through an appropriate model, for instance the multi-species coalescent.
The species tree itself is a tree module, just as the gene tree module, but it is simpler in structure.
For instance, the species tree model might be a birth-death process (see Fig.~\ref{fig:module_alternatives}).
A detailed representation of the link between species tree and gene tree is shown in the supplementary material (Fig.~S.4) for the multi-species coalescent \citep{Heled2010}.

This concludes our introduction to tree plates and modular graphs, essential concepts in providing compact and efficient representations of phylogenetic graphical models.
Tree plates capture the variable, stochastic nature of the dependency structure of the model subgraph corresponding to the phylogeny.
They also exploit the repetitive, recursive structure of phylogenetic trees to provide stringent summaries of the essential details.
Modular graphs are essential in providing high-level, compact representations of large and complex phylogenetic models.
They provide a lot of flexibility through the possibility of collapsing and expanding various model components according to the specific needs in a particular situation.
Large sets of complex models with minor variations in some model components are summarized very efficiently.

\levelone{Computations on Model Graphs}

Probabilistic graphical models, often denoted as Bayesian networks, have long been a major focus in statistics and computer science, and the resulting body of knowledge applies directly to phylogenetic graphical models (PhyloGMs).
In fact, unbeknownst to many in our field, the algorithms used in phylogenetics usually have well-studied equivalents in the computer science literature.
Below, we first provide a mathematical definition of directed acyclic graphs (DAG), and discuss the rationale for using them in PhyloGMs.
We then describe some of the standard algorithms on probability graphs and their applications to phylogenetic problems.
For a more thorough introduction to the field of graphical model algorithms, we direct the reader elsewhere \citep[e.g.][]{Koller2009}.

\leveltwo{Directed Acyclic Graphs (DAGs)}

A directed graph $\mathcal{G}$ consists of a set of nodes (vertices) $\mathcal{V}$ and a set of directed edges $\mathcal{E}$ connecting those nodes, that is, $\mathcal{G} = (\mathcal{V},\mathcal{E})$.
A directed edge from node $a$ to node $b$ is denoted by the pair $(a,b)$.
Direction implies that if $(a,b) \in \mathcal{E}$ then $(b,a) \notin \mathcal{E}$.
A path through the graph is a sequence of nodes, where each node (except the last one) is connected by a directed edge from itself to its successor.
If a path visits the same node twice, the path contains a cycle.
By definition, a directed graph is acyclic if there does not exist any path in the graph that contains a cycle.

DAGs predominate phylogenetic models for two reasons.
First, the relationships among study taxa, on which we build the core of a phylogenetic model, are inherently directed (tipwards) and acyclic because the transmission of genetic material is exclusively from ancestor to descendant.
Second, there are good reasons also from a statistical perspective to focus on DAGs.
A random variable depends on the parameters of its distribution, which form its parents in the model graph.
This is naturally related to causation, and justifies the use of directionality in model graphs.
Undirected or cyclic graphs can be used as well to represent a model, but these representations are complex and typically avoided by statisticians, and currently we see no need for them in phylogenetics.

\leveltwo{Factorization}

The fundamental justification for a graphical model is that it helps us answer questions about the random variables in the model.
Perhaps the most important question concerns the joint probability of a set of variables.
The model graph allows us to compute this efficiently using factorization.
Let us define the set $\mathcal{U}$ as the collection of random variable nodes in the model (with $\mathcal{U}$ being a subset of $\mathcal{V}$).
$\mathcal{U}$ is the complete set of stochastic variables in $\mathcal{V}$ and all remaining variables are either constant or deterministic.
For each $u \in \mathcal{U}$, there is a corresponding random variable in the model, $X_u$.
The set of parent nodes of a node $u$ is denoted by $\pi_u$.
Note, $\pi_u$ denotes all parents in the model graph and not only the single parent specified by tree structure mapping function $p(u)$.
If a variable is indexed by a set of indices such as $\pi_u$ we mean the set of random variables with $\{X_p: p \in\pi_u \}$ and use the short form $\{ X_{\pi_u} \}$. 
Let $x_u$ represent a realization of $X_u$.
For notational convenience, we will assume in this section that all random variables are discrete, although generalization to continuous variables is trivial in most cases (excluding only marginalization and variable elimination).
The conditional independence structure of the model graph allows us to break the problem into pieces (factors), each restricted to one node and its immediate parents, resulting in convenient and efficient computation.
Specifically, given the set of conditional probabilities (or probability density functions) $\{\mathbb{P}(x_u|x_{\pi_u})\}$, the joint probability (density) is obtained as
\begin{equation}
 \mathbb{P}(\{x_u: u \in \mathcal{U}\}) = \prod_{u\in \mathcal{U}}\mathbb{P}(x_u|x_{\pi_u}) \mbox{ .}
\end{equation}

We return to example provided in Figure~\ref{fig:EvolBaculum}.
The model contains the variables with their probability distributions:
\begin{eqnarray*}
p & \sim & \text{Beta}(\alpha=1,\beta=1) \\
\theta & \sim & \text{Beta}(\alpha=1,\beta=1) \\
S_9 & \sim & \text{Bernoulli}(p) \\
S_i & \sim & \text{CTMC}(S_{\tilde{p}(i)},l_i,p) \mbox{ for } i \mbox{ in } \{1,\ldots,8\}
\end{eqnarray*}
Then, the joint probability density of the all variables is
\begin{eqnarray}
\mathbb{P} (p,\theta,S_1,\ldots,S_9) & = & \mathbb{P}_{\text{Beta}}(p,1,1) \times \mathbb{P}_{\text{Beta}}(\theta,1,1) \times \mathbb{P}_{\text{Bernoulli}}(S_9,p) \times \prod_{i=1}^{8} \mathbb{P}_{\text{CTMC}}(S_i,S_{\tilde{p}(i)},l_i) \nonumber \\
& = & p^{S_9}(1-p)^{1-S_9} \times \prod_{i=1}^{8}
\begin{cases}
\theta'_{i} + (1 - \theta'_{i}) \exp(-l_i / (2\theta - \theta^2)) & \mbox{ if } S_i = S_j\\
(1 - \theta'_{i}) * (1 - \exp(-l_i / (2\theta - \theta^2)) & \mbox{ if } S_i \neq S_j
\end{cases} \label{eq:jointDensity}
\end{eqnarray}
where $\theta_i' = \theta$ if $S_i = 1$ and $\theta_i' = 1-\theta$ if $S_i = 0$. This joint probability density can be used to estimate the maximum likelihood parameter estimates, or, as we did in our analysis, to compute the posterior probability density of individual parameters. Equation~(\ref{eq:jointDensity}) is often denoted as the posterior probability density in Bayesian analyses and the posterior density of single parameters is obtained by marginalizing over all other parameters.

\leveltwo{Conditional and Marginal Distribution}

A common set of questions concerns the conditional probability or marginal distribution of one or more random variables (the query nodes), given fixed values of some other variables (the evidence nodes), summarizing over all possible values of (marginalizing out or eliminating) the remaining variables.
For instance, we might have observed the character states of the tip nodes in a phylogenetic tree (evidence nodes), and want to infer the probabilities of the different states of a named interior node (query node), summarizing over all possible state assignments to other interior nodes (remaining nodes).
Formally, let $E$ be the set of (indices of) evidence nodes, $F$ the query node, and $R$ the remaining stochastic nodes.
To obtain the conditional probability of a state $x_F$ of the query node (conditioned only on $x_E$), we need to sum the probabilities over all possible assignments of states to the $R$ nodes.
To obtain the marginal distribution of the query node and the evidence nodes, we need to compute
\begin{equation}
\mathbb{P}(x_E,x_F) = \sum_{x_R} \mathbb{P}(x_E,x_F,x_R) \mbox{,}
\end{equation}
which can be further marginalized over the query node states to give the marginal probability of the evidence nodes
\begin{equation}
\mathbb{P}(x_E) = \sum_{x_F} \mathbb{P}(x_E,x_F) \mbox{,}
\end{equation}
from which we obtain the conditional probability of the query node
\begin{equation}
\mathbb{P}(x_F|x_E) = \frac{\mathbb{P}(x_E,x_F)}{\mathbb{P}(x_E)} \mbox{.}
\end{equation}

The problem here is that $\sum_{x_R}$ expands into a series of summations with a large number of terms.
If there are $|R|$ random variables, each of which can take on $k$ values (\EG four nucleotide states or 20 amino acid states), we have $k^{|R|}$ terms in total.
The large number of terms makes naive summation impossible except in the most trivial cases of very few variables with few states.
The solution is to eliminate the $R$ nodes one by one using the \emph{variable elimination} algorithm \citep{Koller2009}.
The computational complexity of variable elimination depends on the elimination order and the dependency structure of the graph.
In general, finding the optimal order is NP-hard, but good heuristic algorithms are available for the general case, and optimal orderings are known for many common types of graphs such as chain graphs and tree graphs.
Variable elimination algorithms are routinely used for marginal ancestral state reconstruction on phylogenetic trees, and were proposed by \cite{Yang1995b}.

\leveltwo{Sum-Product Algorithm and Belief Propagation}

Trees are important types of graphs, and variable elimination in such graphs is accomplished by the so called \emph{sum-product algorithm} \citep{Gallager1962, Pearl1982, Jordan2004,Ahmadi2012}.
In phylogenetics, the algorithm is known as Felsenstein's pruning algorithm \citep{Felsenstein1981}.
The sum-product algorithm is more limited than variable elimination, in that it is restricted to tree graphs. However, it is more general in that it can compute the marginals of all nodes in the tree using just two passes over the tree, each with the same time complexity as simple variable elimination.
The sum-product algorithm is often described as message passing or \emph{belief propagation}, both important concepts in graphical model algorithms.
Here we provide a short description of belief propagation and refer the reader to \cite{Kschischang2001} and \cite{Ahmadi2012} for more detailed elaborations.

Belief propagation gets its name from the exchange of requests for messages and messages between nodes of the model. In the first pass, requests are propagated, and then in the second pass messages are propagated. More precisely, the algorithm works as follows: 
\begin{enumerate}
\item Send message requests to all neighboring nodes, starting by the (arbitrarily chosen) root node.
\item Only process a request for a message from a neighbor if all messages from other neighbors have been received, and sent out request if necessary.
\item When all messages have been received, compute the marginal probabilities.
\end{enumerate}

A message consists of a vector of (typically unnormalized) marginal probabilities, one for each possible state. For instance, in a nucleotide model there would be four probabilities in the message, one for each state (A, C, G or T). More formally, a node $j$ would send a message $m_{ji}$ to a neighbor $i$ consisting of elements of the kind
\begin{equation}
m_{ji}(x_i) = \sum_{x_j} \left( \mathbb{P}( x_i, x_j ) \prod_{k \in N(j) \setminus i} m_{kj}(x_j) \right) \mbox{,}
\end{equation}
where $N(j)$ are the immediate neighbors of node $j$ in the tree graph, and $\mathbb{P}( x_i, x_j )$ represents the probability of a substitution from state $x_i$ to state $x_j$ (or in the other direction, depending on the direction of the edge connecting $i$ and $j$). The sum and product signs appearing in the message equation give the algorithm its name.

\begin{figure}[!htbp]
\centering
\includegraphics[width=\textwidth]{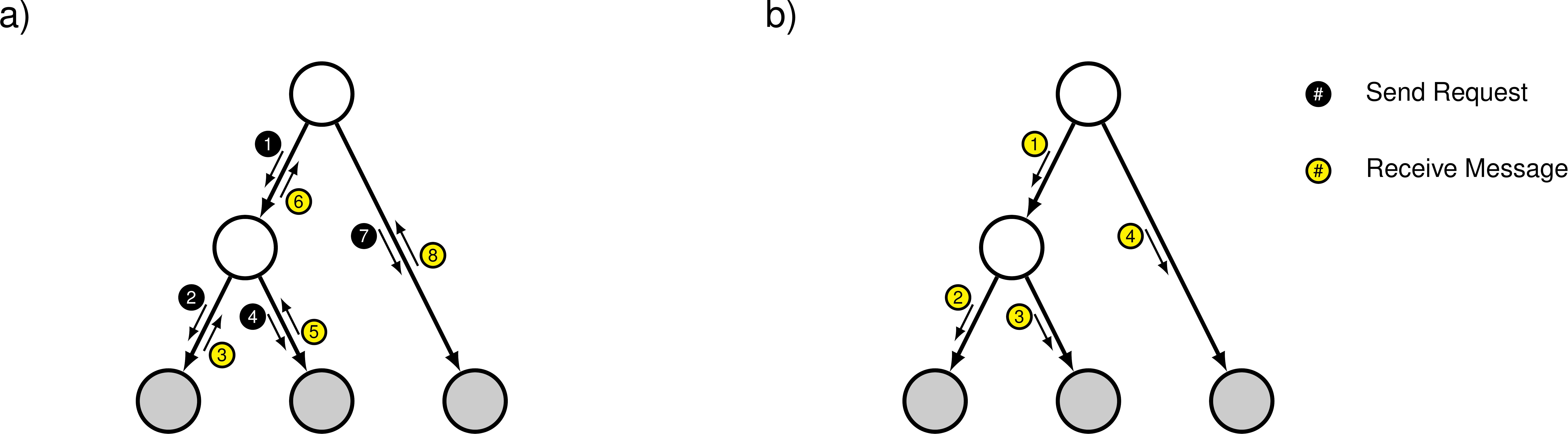} 
\caption{Message passing (belief propagation) on a tree graph.
a) First phase, passing messages from the tips towards the root.
b) Second phase, passing messages from the root towards the tips.
After the second phase, all nodes have received messages from all of their neighbors, and their marginals can be computed.
If only the probability of the entire tree or the marginals of the root node are of interest, the second phase is not needed. }
\label{fig:SumProduct}
\end{figure}

The nodes are visited first in a depth-first (postorder) traversal of the tree to guarantee effective sequential processing, from the tips towards the root, and then in the reverse order (preorder), proceeding from the root towards the tips (Fig.~\ref{fig:SumProduct}).
Undirected (unrooted) trees are rooted first on an arbitrarily chosen node in order to apply the standard traversal algorithms.
When the root has been reached in the first pass, we have all the necessary information to compute the probability (or likelihood) of the whole tree.
We simply multiply all messages received by the root node to obtain the marginal probability for each state. 
Averaging over states then gives $p_E$, the probability of the entire tree given the tip states (the evidence).
Then, the second pass over the tree starts from the chosen root node again and consists only of sent messages of the marginal probabilities for each state towards the tips (see Fig.~\ref{fig:SumProduct}b).
The second pre-order traversal of the tree is only needed if the marginal probabilities are to be computed for other nodes in the tree, for instance, if one wants to draw ancestral states of non-root nodes from the corresponding marginal distributions.

\levelthree{Factor graphs}

Many algorithms on graphical models, such as belief propagation, are designed and/or optimized for factor graphs \citep{Kschischang2001,Loeliger2004,Ahmadi2012}.
Moreover, algorithms studied for various types of graphical models are unified by factor graphs and many general results and insights can thus be transfered from one application to another.
Factor graphs are favored to describe belief propagation because the messages are passed to and from the factor (computation) nodes along the edges containing the variables.

\begin{figure}[!htbp]
\centering
\includegraphics[width=\textwidth]{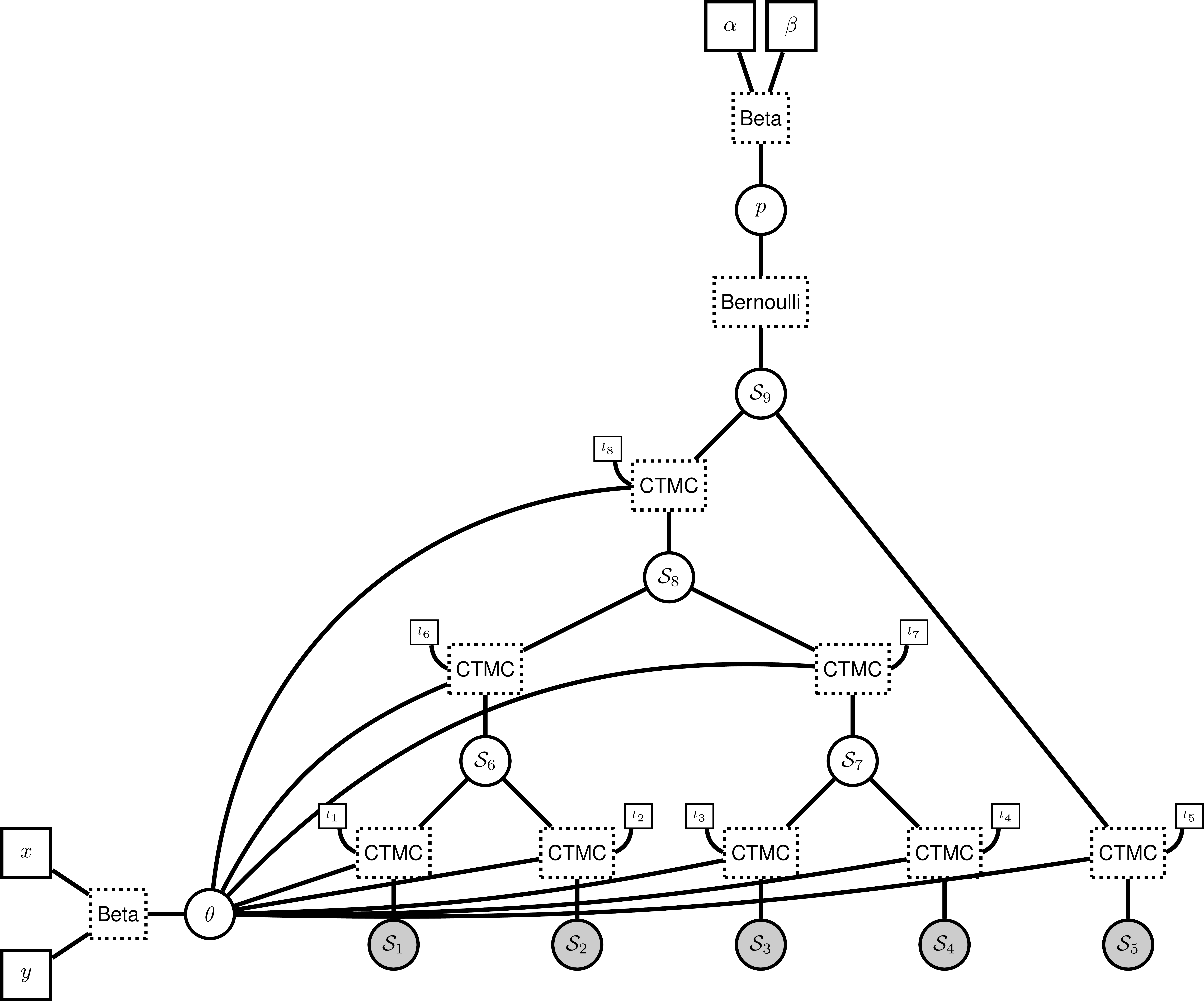}
\caption{
A factor graph representing the binary character evolution model introduced in Figure~\ref{fig:EvolBaculum}. 
The factor graph additionally displays the probability distributions (the factors) as part of the model graph, \EG a Beta distribution, Bernoulli distribution and continuous time Markov chain (CTMC).
A factor graph is always an undirected graph showing only the relationship between the variables and the corresponding distributions.
}
\label{fig:FactorGraph}
\end{figure}

Factor graphs are more fine-grained versions of graphical models, in which the probability distribution (the factors) of each random variable are made explicit by including the distribution as separate nodes in the graph (Fig.~\ref{fig:FactorGraph}).
Furthermore, the direction is dropped in the model graph to show that the computed value of the factor (the probability) depends on the parameters as well as the random value.
Every model graph that is represented by a DAG can be converted into a factor (see \cite{Ahmadi2012} for some examples and elaborations).
We show an example of the conversion in Figure~\ref{fig:FactorGraph}.

The factors, or local functions, are simply the conditional probability density function \citep{Ahmadi2012}.
In the example given in Figure~\ref{fig:FactorGraph} the factorization yields
\begin{eqnarray}
f(p,\theta,S_1,\ldots,S_9) & = & f_{\text{Beta}}(p,\alpha,\beta) \times f_{\text{Beta}}(\theta,x,y) \times f_{\text{Bernoulli}}(S_9,p) \times \prod_{i=1}^{8} f_{\text{CTMC}}(S_i,S_{\tilde{p}(i)},l_i) \nonumber \\ \label{eq:factorGraph}
\end{eqnarray}
which corresponds exactly to Equation~(\ref{eq:jointDensity}).
However, the reverse transformation is not that simple and not every factor graph can be represented as a directed acyclic graph without major modifications.

Belief propagation on factor graphs goes far beyond tree-like (cycle free) graphs and therefore also far beyond Felsenstein's pruning algorithm, which corresponds to the first pass of the algorithm.
It can be extended to accommodate other types of graphs than trees \citep{Loeliger2004}.
A phylogenetic example is the variable elimination in a GTR + $\Gamma$ model, which involves elimination of both character states and rate categories in a graph that is not a tree.
Thus, any additional mixture model component of the substitution process may be integrated/summed over numerically by applying the belief propagation algorithm. 
Hence, belief propagation can be used in various other examples such as a mixture over the rates of positive selection \citep{Yang2002,Huelsenbeck2004b}, mixture over tree topologies \citep{Boussau2009} and mixture over branch rates \citep{Heath2012}.

Modifications of the computation of the message in the belief propagation algorithm can be used to find the maximum a posteriori probability (the so-called Viterbi algorithm \cite{ForneyJr1973}) or the maximum a posteriori configuration over a set of stochastic nodes (max-product or min-sum algorithm \cite{Tanner1981}).
An example of the latter would be the computation of the set of character states at ancestral nodes most likely to have produced an observed set of tip states.
Belief propagation is a type of dynamic programming, which is one of the most important techniques in computational optimization.

\leveltwo{MCMC Sampling}

Markov chain Monte Carlo (MCMC) sampling is a core technique used in Bayesian inference.
It is relatively straightforward to set up a Markov chain that has the distribution of interest, the posterior probability, as its stationary distribution but convergence to the target distribution is often relatively slow.
Therefore, the algorithm needs to be run for many generations, and computational efficiency is paramount.
Model graphs provide an elegant way of structuring the conditional dependencies in such a way that the computational efficiency of MCMC algorithms can be maximized.
It is no coincidence that BUGS \citep{Spiegelhalter1990, Lunn2000, Lunn2009,Lunn2012}, one of the most successful software packages for Bayesian inference, is built entirely around graphical models.
In fact, the BUGS team were among the early adopters of graphical models and contributed importantly to their development, \EG by introducing deterministic nodes to capture variable transformations.

We illustrate the use of graphical models in Bayesian MCMC sampling in the context of the standard Metropolis-Hastings algorithm \citep{Metropolis1953,Hastings1970}. In iteration $t$ of the algorithm, the stochastic nodes $\mathcal{U}$ start out having the values $x^{(t)} = \{ x_u^{(t)} \}$. The iteration then consists of the following steps:
\begin{enumerate}
\item Propose new values $x'$ according to a proposal density $q(x' |x^{(t)})$.
\item Compute the acceptance probability
$
\alpha = \min\left(1,\frac{\mathbb{P}(x')}{\mathbb{P}(x^{(t)})} \times \frac{q(x^{(t)} | x')}{q(x' | x^{(t)})}\right)
$.
\item With probability $\alpha$ accept the proposal and set $x^{(t+1)} = x'$; otherwise reject the proposal and set $x^{(t+1)} = x^{(t)}$.
\end{enumerate}

The computationally expensive step is to obtain the ratio of the joint probability of the model before and after the proposal, $p(x') / p(x^{(t)})$.
In theory, a proposal could involve changing values of all non-clamped stochastic nodes in the model, making it difficult to achieve computational efficiency.
In practice, however, a mixture of many different proposal mechanisms is used, with each proposal changing the value of only one or a few stochastic nodes. 
Taking advantage of the conditional-independence factorization provided by the graphical model formalism, we can quickly identify the minimal set of conditional probabilities that need to be updated.

Consider a proposal changing just one stochastic node $i$ and let $c(i)$ denote the children of that node.
In principle, we need to calculate
\begin{equation*}
\frac{\mathbb{P}(x')}{\mathbb{P}(x^{(t)})} = \prod_{u \in \mathcal{U}} \frac{ \mathbb{P}( x_u' | x_{\pi_u}' ) }{ \mathbb{P}( x_u^{(t)} | x_{\pi_u}^{(t)} ) }\mbox{,}
\end{equation*}
a product over all nodes in the graph.
However, for all nodes in $\mathcal{U}$ except $i$ and $c(i)$, the conditional probabilities are going to be the same before and after the move.
Therefore, we can simplify the calculation to
\begin{equation*}
\frac{\mathbb{P}(x')}{\mathbb{P}(x^{(t)})} = \frac{ \mathbb{P}( x_i' | x_{\pi_i}^{(t)} ) } { \mathbb{P}( x_i | x_{\pi_i}^{(t)} ) } \times
           \prod_{u \in c(i)} \frac{ \mathbb{P}( x_u^{(t)} | x_i' ) } { \mathbb{P}( x_u^{(t)} | x_i^{(t)} ) } \mbox{.}
\end{equation*}
Thus, only the changed node and its children need to be considered in calculating the model probability ratio \citep{Spiegelhalter1990}.
Similarly, if the proposal changes the values of a set of nodes rather than a single node, it is sufficient to consider the changed nodes and their children in calculating the probability ratio.
As an illustrative example consider the case when a new value for the probability $p$ of a baculum of the common ancestor of all taxa is proposed (see Figure~\ref{fig:EvolBaculum}).
The joint probability density was given in Equation~(\ref{eq:jointDensity}) and the computation for the full data set contains many factors.
However, the probability ratio simplifies to
\begin{eqnarray}
\frac{\mathbb{P}(p')}{\mathbb{P}(p^{(t)})} & = & \left(\frac{p'}{p^{(t)}}\right)^{S_9} \left(\frac{1-p'}{1-p^{(t)}}\right)^{1-S_9} \end{eqnarray}
regardless of how many taxa are included in the study. 
The probability ratio is clearly simpler than the joint posterior probability density and the ratio thereof and the computation is much faster.

Finally, consider Gibbs sampling \citep{Geman1984,Gelfand1990}, a special case of the Metropolis-Hastings algorithm, in which the proposal distribution is the posterior probability distribution of the changed variable(s), conditional on the values of the other random variables in the model.
Simultaneous Gibbs sampling of all unclamped random variables in a model would be equivalent to random draws from the target distribution, which is difficult to beat in terms of sampling performance.
In practice, one is happy if it is possible to do Gibbs sampling of individual random variables in the model.
Specifically, Gibbs sampling of a random variable is possible when the distribution from which it is drawn (the prior) is conjugate with respect to its conditional posterior.
In this context, conjugate means that the two distributions come from the same family of distributions.
The graphical model structure is helpful both in checking for conjugate distributions and in implementing Gibbs sampling where it is feasible.
This property of graphical model has been exploited extensively by BUGS \citep{Spiegelhalter1990, Lunn2000, Lunn2009,Lunn2012}.

\leveltwo{Simulation}

Simulating data from a model is essential in many applications, for instance in exploring model properties \citep{Huelsenbeck1995} or in model adequacy testing \citep{Bollback2002,Brown2009,Hohna2013a}.
Simulations are also used to validate inference methods and to initialize MCMC runs.
One way of simulating from a model is to use MCMC sampling.
In fact, ordinary MCMC sampling can be understood as a way of simulating draws from the unclamped random variables of the model.
By simply changing all clamped nodes to unclamped ones, we can generate (dependent) samples from the full model.
However, MCMC sampling is not necessarily the most efficient simulation strategy.
Completely independent samples from the model can be generated by simply traversing the model graph from the source nodes towards the sink nodes, drawing values of each random variable conditioned on the already generated values of its parent nodes.
In both cases, the model graph formalism provides a natural infrastructure.

\leveltwo{More Computation on Model Graphs}
In this section, we have only skimmed the surface of the literature on graphical-model computation.
We have not mentioned methods that allow efficient maximum likelihood inference by using the structure of graphical models, such as the expectation maximization (EM) algorithm or variational methods that minimize Kullback-Leibler divergence \citep{Koller2009}.
We have not discussed analysis of conditional independence, and many other methods of interest.
However, our examples have hopefully demonstrated the relevance of the rich graphical-model literature to statistical phylogenetics.
Our point is not that phylogeneticists have necessarily been hampered significantly thus far by ignoring graphical models.
However, the benefits of adopting the graphical models framework will increase rapidly over the coming years, as phylogenetic models become increasingly complex.
However, the benefits of adopting the graphical models framework will increase rapidly over the coming years and as phylogenetic models become increasingly complex by facilitating better communication with the statistics community.


\levelone{Discussion}

Statistical phylogenetics has developed to the point where the number and complexity of phylogenetic models are posing serious challenges to theoreticians, empiricists and software developers alike.
It would represent a big step forward if the field adopted a standardized and efficient way of describing phylogenetic models and exposing their underlying structure.
We argue that the graphical models framework, used by statisticians to address similar challenges, provides an appropriate tool to this end.
Graphical models have not been used in phylogenetics previously but they have been applied to many other research areas and several workers have suggested their use in phylogenetics \citep{Lunn2000,Friedman2002,Friedman2004,Jordan2004,Lunn2009,Koller2009}.

Graphical models are based on the idea of breaking large probabilistic models into components representing conditionally independent probability distributions.
Additional representational power is obtained by using plates for replication and deterministic nodes for variable transformations.
Although many aspects of phylogenetic models can be readily described using these standard graphical model concepts, the phylogenetic models also present some special difficulties.

The core part of a PhyloGM, the one corresponding to the evolutionary tree, is unusual in a graphical models perspective both because it can be so large and because the graph structure (the topology) is considered a random variable subject to estimation.
To address these challenges, we introduced tree plates.
They allow efficient representation of large trees with many tips and they also capture the structure learning nature of tree topology inference.
We further simplified the representation of large and complex phylogenetic models by introducing a modular representation that breaks them into connected subgraphs at carefully chosen variable nodes, called pivot nodes.
The modular representation is highly flexible, allowing both compact high-level representation of models and efficient detailed exposition of the model subgraphs of particular interest.
By combining different modules in various patterns, a large set of models can be represented very efficiently.

With the addition of tree plates and modularization, we believe that graphical models are ready for wide use in the statistical phylogenetics community.
They provide a rich framework for teaching and communicating probabilistic models.
With their explicit representation of assumptions and variable dependencies, they facilitate the understanding of complex models and they reduce the risk of similar models being confused.
Graphical models should be useful both for empiricists who want to learn the essential features of models and for theoreticians who want to communicate new models and put them in the context of previously published models.

Adopting the graphical model approach would also help connect statistical phylogenetics to other science areas, promoting interdisciplinary cross-fertilization that may well turn out to be productive.
For example, graphical models have been well studied from a computational perspective.
Many algorithms are known for efficiently computing joint or marginal probabilities, and for performing MCMC sampling or simulation on probabilistic model graphs.
In fact, as we have shown, many of the standard algorithms used in computational phylogenetics have older and well-studied equivalents in the literature on model graphs.
As phylogenetic models grow in complexity in the future, the existing work on model graph algorithms may well prove to be a treasure trove for phylogeneticists, greatly facilitating the development and implementation of new models.

Graphical models may also help forge links between statistical phylogenetics and other fields of applied statistics. 
Applied statisticians often summarize models using formulae of the type $y \sim f(\alpha,\beta)$, specifying that a random variable $y$ is drawn from some distribution $f$ with parameters $\alpha$ and $\beta$ \citep[for a range of examples, see][]{Lunn2012}.
Such model formulae are rarely used in phylogenetics today.
However, they are closely related to graphical model concepts, so phylogeneticists adopting this framework are likely to find such model formulae helpful and informative summaries of their models.
This, in turn, will make it easier for applied statisticians to contribute to phylogenetics.

Last but not least, the adoption of graphical models would facilitate the design and development of computational phylogenetics software.
There are decidedly some challenges involved in doing this, particularly in finding efficient software representation of the huge PhyloGMs.
However, regular plates and tree plates help identify some of the replicated structure that can be used in efficient implementation of PhyloGMs.
Modularization also encourages good software engineering principles, in that it supports a natural, high-level design with exchangeable and re-usable components corresponding to standard modules in PhyloGMs.
We end the paper by briefly presenting a software implementation illustrating some of these points.

\leveltwo{Software Implementation}

BUGS (Bayesian inference using Gibbs sampling) is the dominant software package for Bayesian inference \citep{Lunn2000,Sturtz2005,Lunn2009,Lunn2012}.
It defines its own modeling language, the BUGS language, which is entirely based on graphical model concepts. 
A model is specified by setting up the dependency structure of the variables, both deterministic and stochastic, in a special model definition file.
This file is compiled into a model graph, and once the data and initial values are read in, the posterior probability distribution can be estimated using Gibbs sampling or, more recently, the Metropolis-Hastings algorithm.
The focus in BUGS is on linear models, even though a few other model types are also available.

Unfortunately, BUGS is not suited for PhyloGMs.
PhyloGMs include a number of variable types and probability distributions that are not implemented in BUGS, such as tree topologies, instantaneous rate matrices, and continuous time Markov chains.
The domain-specific PhyloGM objects also put special demands on the computational machinery, such as efficient belief propagation in tree plates and effective MCMC samplers of tree topologies \citep{Lakner2008,Hohna2012}.
Furthermore, most PhyloGMs include a graph learning problem, in that part of the graph structure (the topology of the phylogenetic tree) is considered a random variable.
Currently, such inference problems are foreign to BUGS.
Finally, PhyloGMs are considerably larger than most other graphical models, raising significant challenges in handling the objects in a manner that allows fast computation and leaves a small memory footprint.

The limitations of BUGS motivate an independent software implementation for PhyloGMs.
We provide such an implementation in \emph{RevBayes} (www.RevBayes.net).
The software will be presented in more detail elsewhere but we briefly outline it here.
RevBayes provides a command-line interface for interactive analyses, much like the widely used statistical software package \emph{R} \citep{RCoreTeam2013}.
However, unlike R and BUGS, RevBayes allows users to interactively construct complex graphical models, step by step, and it supports all the objects needed to build PhyloGMs.
Like BUGS, the specification of a model closely mirrors its visual graphical-model representation.
The language used by RevBayes, \emph{Rev}, combines features of the R and BUGS languages with those of popular object-oriented programming languages.
The similarities in syntax between Rev and R are intended to help users with previous R experience to learn the language quickly.

We show a simple example here of a Rev model specification. It is equivalent to the binary character model of Figure~\ref{fig:binaryFig1}:
\begin{verbatim}
# Create a vector of observations
data <- {1,1,1,0,0}

# Create constant variables named 'a' and 'b'
# and assign the value 1 to them
a <- 1
b <- 1

# Create a stochastic variable drawn from
# a beta distribution
p ~ beta( a, b )

# Create a vector of stochastic nodes drawn
# from a Bernouilli distribution and clamp 
# them to the data
for (i in 1:5) {
    x[i] ~ bernoulli( p )
    x[i].clamp( data[i] )
}
\end{verbatim}
The other examples used in this manuscript are provided as Rev files in the supplementary material.

\newpage
{\renewcommand{\bibsection}{\levelone{References}}

\bibliography{literature}


\newpage
\renewcommand\thefigure{S.\arabic{figure}}
\setcounter{figure}{0}

\begin{flushright}
Version dated: \today
\end{flushright}
\bigskip
\noindent RH: GRAPHICAL MODELS IN PHYLOGENETICS

\begin{center}

\noindent{\Large \bf Probabilistic Graphical Model Representation in Phylogenetics\\ Supplementary information}
\medskip

\noindent {\normalsize \sc Sebastian  H\"ohna$^1$, Tracy A. Heath$^2$, Bastien Boussau$^{2,3}$, Michael J. Landis$^2$, Fredrik Ronquist$^4$ and John P. Huelsenbeck$^2$}\\
\medskip

\noindent {\small \it 
$^1$Department of Mathematics, Stockholm University, Stockholm, SE-106 91 Stockholm, Sweden;\\
$^2$Department of Integrative Biology, University of California, Berkeley, CA, 94720, USA;\\
$^3$Bioinformatics and Evolutionary Genomics, Universit\'e de Lyon, Villeurbanne, France;\\
$^4$Department of Biodiversity Informatics, Swedish Museum of Natural History, SE-10405 Stockholm, Sweden}\\
\end{center}

\medskip
\noindent{\bf Corresponding author:} Sebastian  H\"ohna, Department of Mathematics, Stockholm University, Stockholm, SE-106 91 Stockholm, Sweden; E-mail: sebastian.hoehna@gmail.com.\\

\newpage

\levelone{Additional phylogenetic graphical models}

\leveltwo{A model of continuous character evolution}

\begin{figure}[!htbp]
\centering
\includegraphics[width=5in]{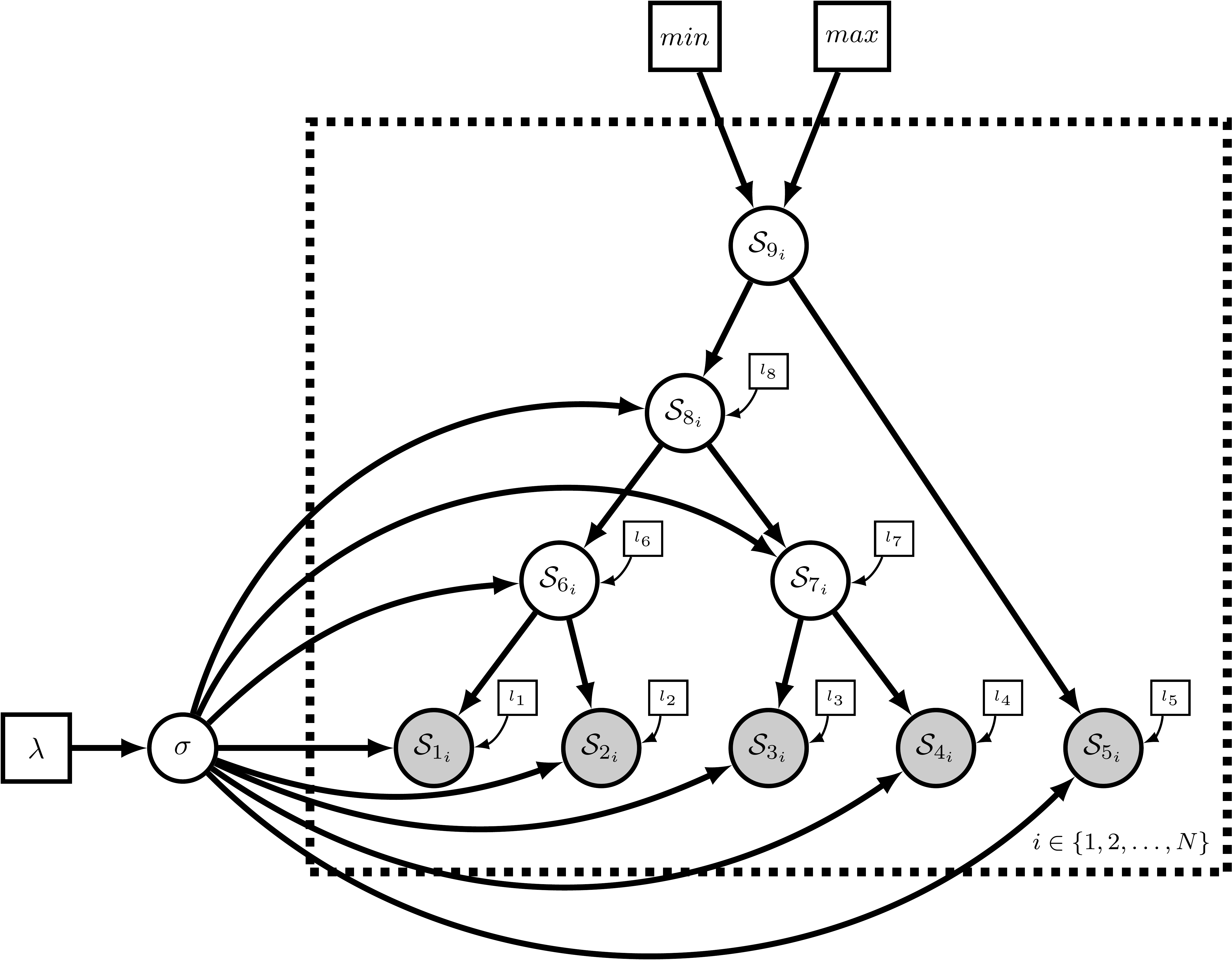} 
\caption{Graphical model of continuous trait evolution example.}
\label{fig:treeContinuous}
\end{figure}

Graphical models are high-level representations that do not depend on details of the model. 
As a result, similar models will have similar representations. 
We provide here the example of a Brownian motion model of the evolution of continuous characters to convey this point \citepsm{Felsenstein1985}. 
We use the same 5 species phylogeny as before. 
Figure~\ref{fig:treeContinuous} shows that the graphical representation of this model is very similar to that of the former binary model, notably because the structure of the phylogenetic tree is still very obvious, and because a plate indicates replication over several characters. 
Only a few details differ between figures Figs.~4 and \ref{fig:treeContinuous}, that describe peculiarities of the model of character evolution. 
In particular, the ancestral state at the root for the Brownian motion model is drawn uniformly between values $min$ and $max$, and the parameter $\delta$ for the variance of the model is drawn uniformly between $x$ and $y$.

\FloatBarrier
\leveltwo{A mixture model for rate variation among sites}

\begin{figure}[!htbp]
\centering
\includegraphics[width=\textwidth]{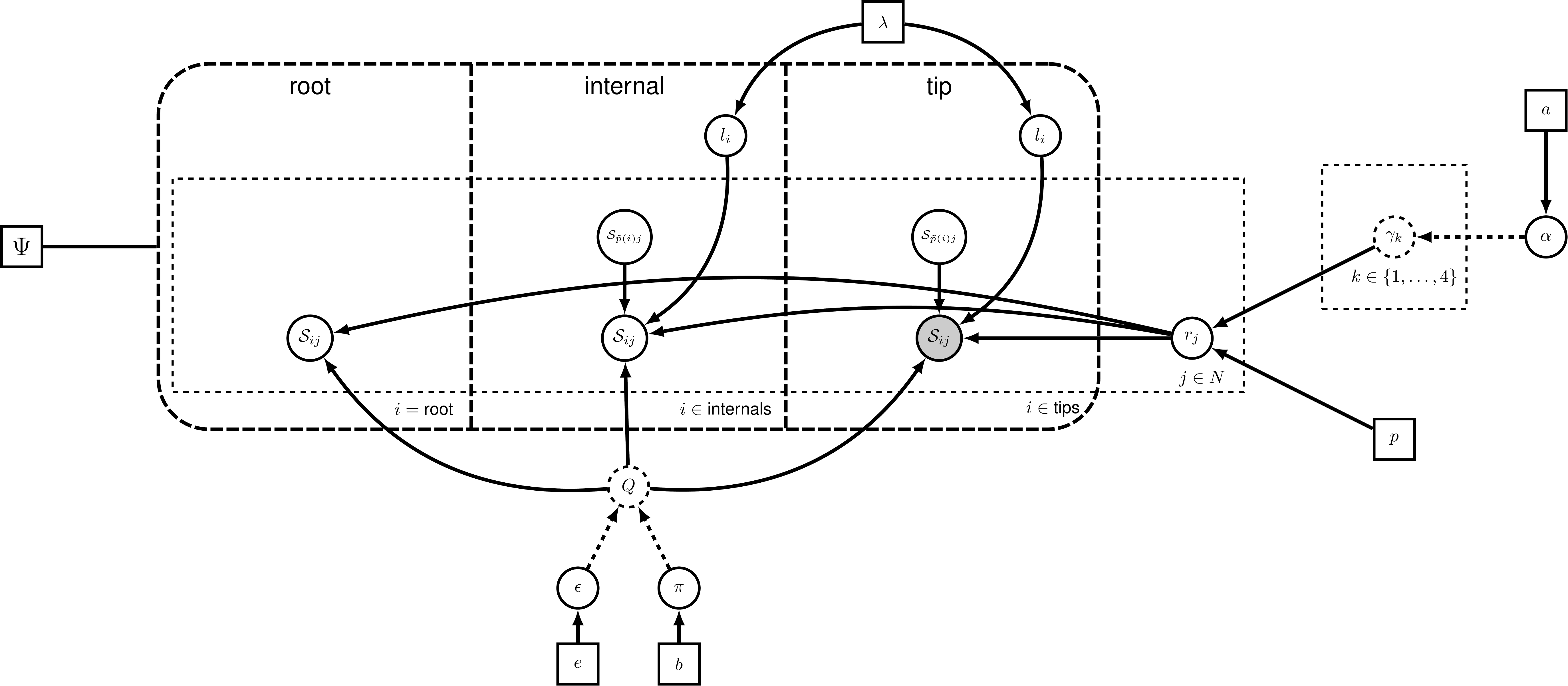} 
\caption{Graphical model representation of among site rate variation model (GTR+$\Gamma$).}
\label{fig:ASRVmodel}
\end{figure}

Mixture models are very common in phylogenetics and here we give an example of a phylogenetic mixture model -- the GTR model with rate variation among sites drawn from four possible rate categories \citepsm{Yang1994,Yang1996} -- represented by a model graph.
The mixture component of this model is the specific rate for each site in the sequence.
Hence, the mixture distribution is modeled by a simple multinomial distribution with four possible outcomes, the four possible rates, and equal probabilities $p$ for each rate category .
Then, every site evolves under one out of the four rates and each rate is computed by the quantiles ($q \in \{0.125, 0.375, 0.625, 0.875\}$) of a gamma distribution with rate parameter $\alpha$ and shape parameter $\alpha$ ($\gamma_i = \text{qgamma}(q_i,\text{shape}\!=\!\alpha,\text{rate}\!=\!\alpha)$).

In general, mixture models can be represented by a multinomial distribution and the category one observation belongs to can be obtained via an indicator variable.
Additionally, the actual mapping to the mixture category can be integrated over by summation of the probabilities of being in each category or marginalized over within the MCMC algorithm.
The graphical model framework does not restrict to any of these methods and can be applied to many types of mixture models, for example infinite mixture models as the Dirichlet Process Prior model \citepsm{Huelsenbeck2006,Heath2012a,Heath2012b}.

\newpage
\leveltwo{A autocorrelated relaxed clock model}
\FloatBarrier

\begin{figure}[!htbp]
\centering
\includegraphics[width=\textwidth]{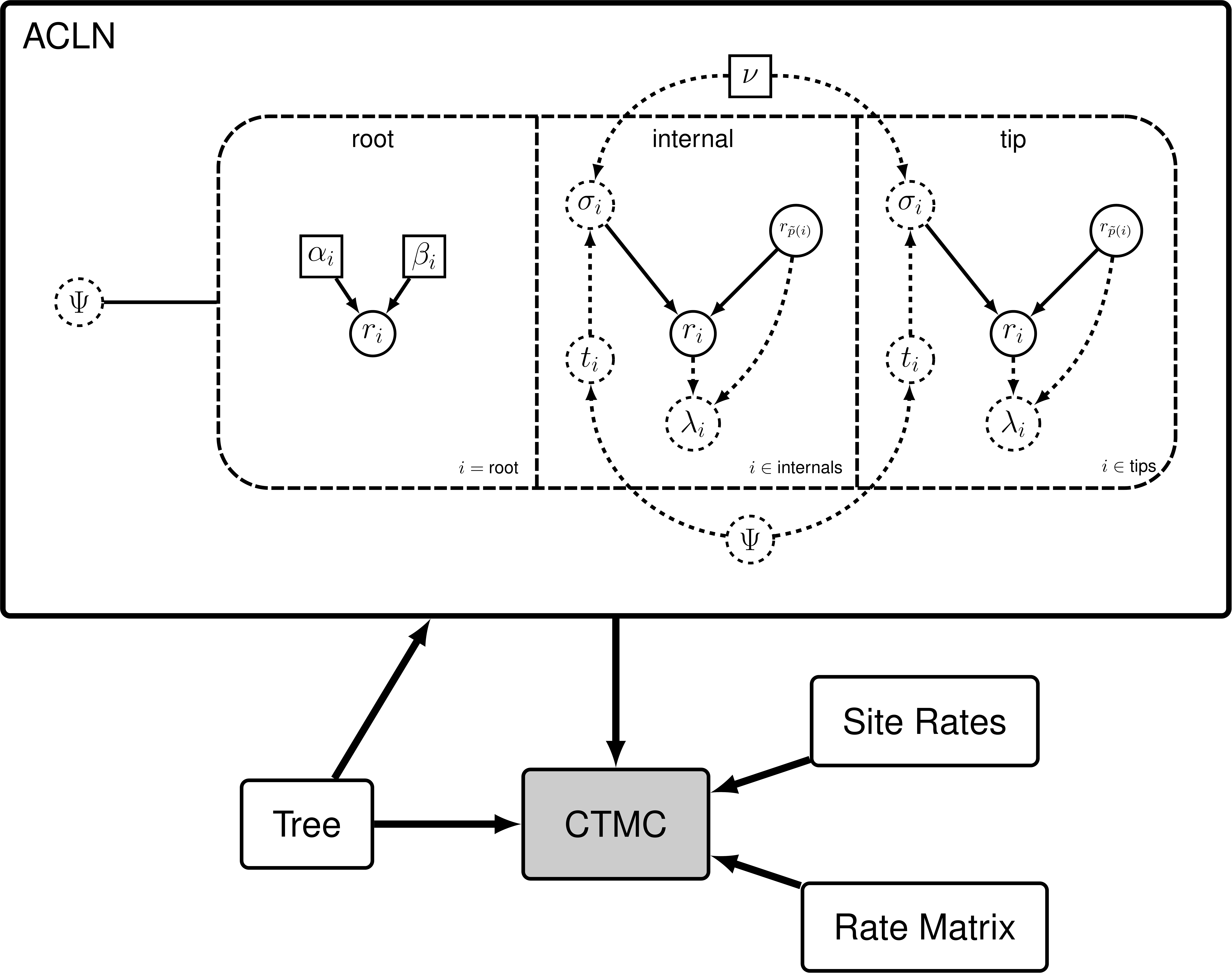} 
\caption{Treeplate GM representation of ACLN}
\label{fig:ACLNTreeplate}
\end{figure}

Relaxed clock models are commonly used when one is interested in dating the divergence between species and here we provide an example of a graphical module for the autocorrelated relaxed clock model of \citesm{Thorne1998}.
The autocorrelated relaxed clock specifies that the logarithm of the clock rates evolves under a Brownian motion ($r_i \sim \text{norm}(\text{mean}\!=\!r_{p(i)},\text{variance}\!=\!t_i\nu)$).
The tree plate enables a compact representation of this relaxed clock but still emphasizes that the clock rates depend on the tree topology.

%
%
%

\newpage
\leveltwo{A graphical multispecies coalescent model}
\FloatBarrier

\begin{figure}[!htbp]
\centering
\includegraphics[width=\textwidth]{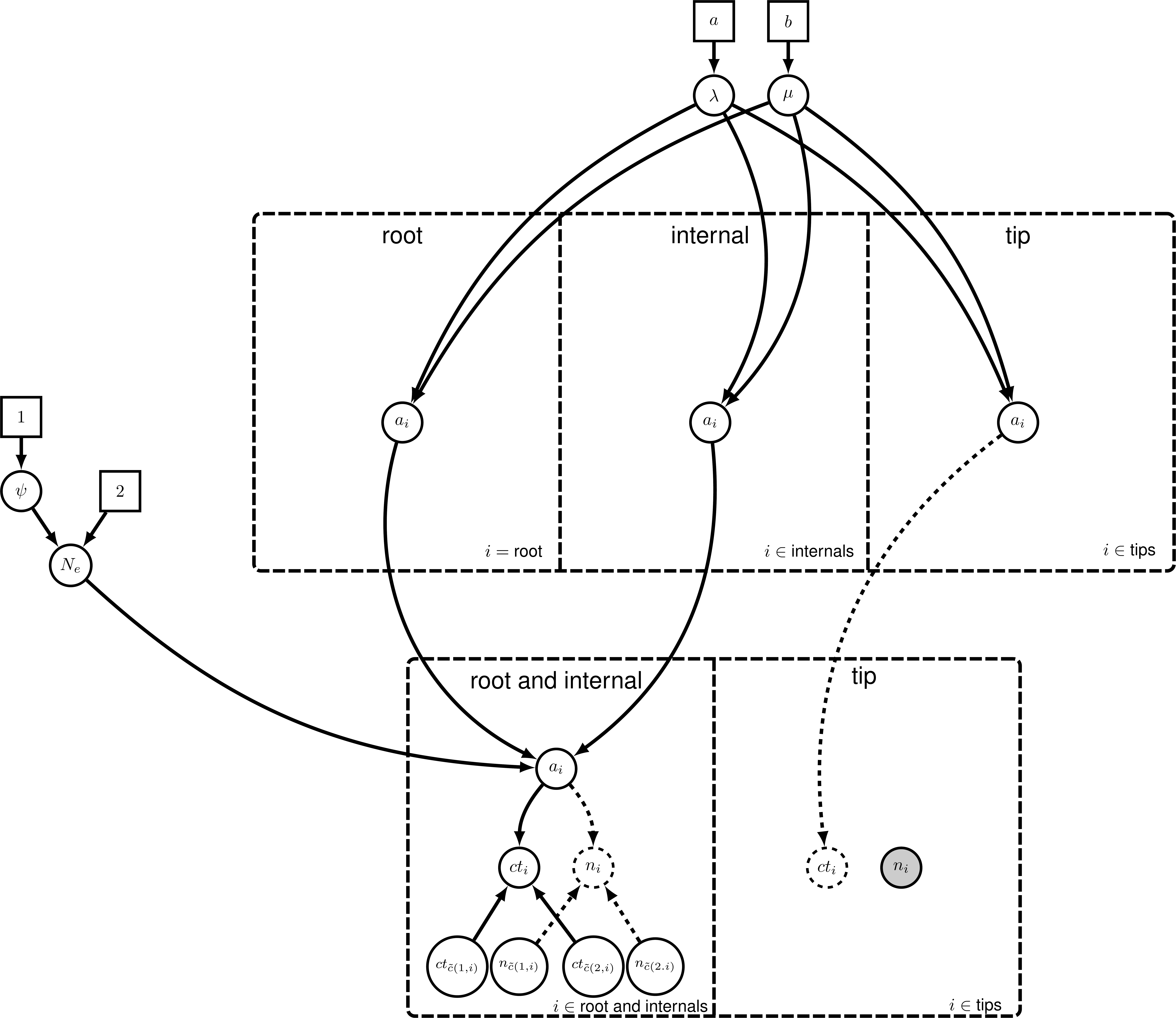} 
\caption{Treeplate GM representation of the multispecies coalescent model.}
\label{fig:MultispeciesCoalescent}
\end{figure}

Here we provide a model graph using tree plates for a multispecies coalescent model.
First, the birth-death process \citepsm{Nee1994b} acts as a prior distribution on the speciation times (or ages $a_i$) of the species tree.
Then, each branch of the species trees represent an independent population and a coalescent process is applied to the coalescent times of the gene trees \citepsm{Rannala2003,Heled2010}.
The dependence between the speciation times of the species tree and the coalescent times in the gene trees is emphasized by this model graph.
For multiple genes one could simply put the gene tree model on a plate.

\newpage
\levelone{RevBayes Examples}

In the following section we have listed the two \textit{bacula} examples, a non-phylogenetic example assuming no dependence structure between species and a phylogenetic example using the phylogenetic tree as structural dependence.

\leveltwo{A non-phylogenetic example}
In this example we have observations on the presence and absence of a baculum for five species.
We model this by a Bernoulli distribution and estimate the parameter $p$.
This example is given in the Rev language and thus can be run in \emph{RevBayes}.
It is available as a runnable file separately too.

\begin{verbatim}
# set the prior parameters
alpha <- 1      # this creates a constant variable with value 1
beta <- 1

# create the stochastic variable for the parameter of the binomial distribution
p ~ beta(alpha,beta)    # this creates a stochastic variable drawn from 
                        # a beta distribution with parameters alpha and beta

# create the data
data <- [1,1,1,0,0]     # this creates a vector

for (i in 1:data.size()) {
  x[i] ~ bernoulli(p)   # this creates a stochastic variable drawn from 
                        # a Bernoulli distribution with parameter p
  # attach/clamp the data
  x[i].clamp(data[i])
}

# create the model from the DAG
mymodel <- model(p)


# create the moves/proposals that change the parameters
# of the model during the MCMC
moves[1] <- mSlide(p, delta=0.2, weight=1.0)

monitors[1] <- modelmonitor(filename= "GraphicalModels_Example_1a.log",
printgen=10, separator = "	")
monitors[2] <- screenmonitor(printgen=10, separator = "	", p)

 
mymcmc <- mcmc(mymodel, monitors, moves)

# If you choose more or different proposals, 
# or different weights for the proposals, 
# then the number of proposals changes per iteration. 
# Currently there is only one proposal with weight 1.0.
mymcmc.burnin(generations=2000,tuningInterval=100)
mymcmc.run(generations=200000)

result <- readTrace("GraphicalModels_Example_1a.log")
result[5]
\end{verbatim}

\newpage
\leveltwo{A phylogenetic example}
The second example includes all 274 mammalian species included by \citesm{Reis2012}.
The data is read in from file.
This examples assumes an underlying phylogenetic model and specifies the evolution of the presence/absence of the baculum by a continuous time Markov model.
Note, that we assume a different root frequency of the baculum than the stationary frequency of the continuous time Markov model (see Fig.~3.b).

\begin{verbatim}
# Read the data.
# The readCharacter function returns a vector of matrices.
# We just take the first one.
D <- readCharacterData("data/baculumData01.nex")[1]

# Get some useful variables from the data
nSites <- D.nchar()[1]

# Create the parameter for the root frequencies. 
# Instead of a beta distribution we use the Dirichlet distribution
# because the data type needs to be a simplex. 
# Set a flat prior.
rf_prior <- [1,1]
# Create the random variable for the root frequencies
rf ~ dirichlet(rf_prior)

# Create a move/proposal that changes the root-frequencies during the MCMC.
moves[1] <- mSimplexElementScale(rf, alpha=10.0, tune=true, weight=2.0)


# Now let us create the random variables for the continuous time Markov process
# that changes the absent/present state along the tree.
# We use the F81 rate matrix again with a Dirichlet distribution
# as the prior on the base-frequencies.
bf_prior <- [1,1]
# Create the random variable for the base-frequencies.
bf ~ dirichlet(bf_prior)
# Construct the rate matrix.
Q := F81(bf)

# Create a move/proposal that changes the base-frequencies during the MCMC.
moves[2] <- mSimplexElementScale(bf, alpha=10.0, tune=true, weight=2.0)


# We use a fixed tree (dos Reis et al.) read from a file.
tau <- readTrees("data/mammalia_dosReis.tree")[1]

# Just use the default clock rate. (We could also omit this parameter.)
clockRate <- 1.0

# Construct a random variable for the sequence evolution model.
seq ~ substModel(tree=tau, Q=Q, branchRates=clockRate, 
                      rootFrequencies=rf, nSites=nSites, type="Standard")

# Attach the data.
seq.clamp(D)


# Create the model from the DAG.
mymodel <- model(Q)


monitors[1] <- modelmonitor(filename= "GraphicalModels_Example_2.log",
                              printgen=10, separator = "	")
monitors[2] <- screenmonitor(printgen=10, separator = "	", bf, rf)

 
mymcmc <- mcmc(mymodel, monitors, moves)

# If you choose more or different proposals, or different weights for the proposals, 
# then the number of proposals changes per iteration. 
# Currently there are only two proposal with weight 2.0 each.
mymcmc.burnin(generations=2000,tuningInterval=100)
mymcmc.run(generations=200000)

result <- readTrace("GraphicalModels_Example_2.log")
result[5]
\end{verbatim}

The estimated root frequency is $\text{rf} = 0.503$ with HPD$= \{0.07,0.91\}$ and the estimated equilibrium frequency is $\text{rf} = 0.48$ with HPD$= \{0.38,0.58\}$.
Although there is a slight difference in the parameter estimates, our assumption of a different root frequency is not supported by the data.

\newpage
{\renewcommand{\bibsection}{\levelone{References}}

\bibliographystylesm{apalike}
\bibliographysm{literature}

\end{document}